# Achievable Rates of Buffer-Aided Full-Duplex Gaussian Relay Channels

Ahmed El Shafie[†], Ahmed Sultan[⋆], Ioannis Krikidis[∗], Naofal Al-Dhahir[†], Ridha Hamila[††]

[†]University of Texas at Dallas, USA.
[⋆]King Abdullah University of Science and Technology (KAUST), Saudi Arabia.
[∗]University of Cyprus, Cyprus.
[††]Qatar University.

*Abstract*—We derive closed-form expressions for the achievable rates of a buffer-aided full-duplex (FD) multiple-input multiple-output (MIMO) Gaussian relay channel. The FD relay still suffers from residual self-interference (RSI) after the application of self-interference mitigation techniques. We investigate both cases of a slow-RSI channel where the RSI is fixed over the entire codeword, and a fast-RSI channel where the RSI changes from one symbol duration to another within the codeword. We show that the RSI can be completely eliminated in the slow-RSI case when the FD relay is equipped with a buffer while the fast RSI cannot be eliminated. For the fixed-rate data transmission scenario, we derive the optimal transmission strategy that should be adopted by the source node and relay node to maximize the system throughput. We verify our analytical findings through simulations.

## I. INTRODUCTION

Relay nodes play an important role in wireless communications due to their ability to increase the data rate between a pair of communicating nodes [1]. Relays can operate in three different modes, namely, half-duplex (HD) mode [2]–[4], full-duplex (FD) mode [5]–[11], or hybrid HD/FD mode [12]–[14]. In the FD mode, data transmission and reception at the FD relay node occur simultaneously and over the same frequency band. However, due to the simultaneous reception and transmission, FD relays are impaired by loopback self-interference (LSI), which occurs due to energy leakage from the transmitter radio-frequency (RF) chain into the receiver RF chain [15]–[18]. LSI can be suppressed by up to 120 dB in certain scenarios, as discussed in [19]. However, the LSI cancellation process is never perfect, thereby leaving some non-negligible residual self-interference (RSI). In many modern communication systems such as WiFi, Bluetooth, and Femtocells, the nodes' transmit power levels and the distances between communicating nodes have been decreasing. In such scenarios, the high computation capabilities of modern terminals enable efficient implementation of the FD radio technology [20]–[22]. In the HD mode, transmission and reception occur over orthogonal time slots or frequency bands. As a result, HD relays do not suffer from RSI, but at the cost of

This paper was made possible by NPRP grant number 8-627-2-260 from the Qatar National Research Fund (a member of Qatar Foundation). The statements made herein are solely the responsibility of the authors.

wasting time and frequency resources. Hence, the achievable data rates of an FD relaying system might be significantly higher than that of an HD relaying system when the RSI has low power. In the hybrid HD/FD mode [14], the relay can operate in either HD mode or FD mode to maximize the achievable rate. The key idea is to dynamically switch between the two modes based on the RSI level. When the RSI level is high, the HD mode can achieve higher rates. On the other hand, when the RSI level is low, the FD mode can result in much higher data rates.

Integrating multiple-input multiple-output (MIMO) techniques with relaying further improves the communication performance and data rates [23], [24]. Although most previous research efforts have focused on MIMO-HD relaying, recent research has also investigated the performance of MIMO-FD relaying [25]–[27]. MIMO techniques provide an effective means to mitigate the RSI effects in the spatial domain [25]–[27]. With multiple transmit or receive antennas at the FD relay node, data precoding at the transmit side and filtering at the receive side can be jointly optimized to mitigate the RSI effects. Minimum mean square error (MMSE) and zero forcing (ZF) are two widely adopted criteria in the literature for the precoding and decoding designs [28]. ZF aims to completely cancel out the undesired self-interference signals and results in an interference-free channel at the relay node's receive side. Although ZF normally results in a sub-optimal solution to the achievable performance (i.e., data rate and bit error rate), its performance is asymptotically optimal in the high signal-to-noise ratio (SNR) regime. On the other hand, MMSE improves the performance of the precoder/decoder design compared to ZF since it takes into account the noise impact at the cost of a higher complexity. However, due to the implementation simplicity and optimality in the high-SNR regime, ZF has been proposed as a useful design criterion to completely cancel the RSI and separate the source-relay and relay-destination channels.

Assuming there is no processing delay at the relay, the optimal precoding matrix for a Gaussian FD amplify-and-forward (AF) relay that maximizes the achievable rate under an average power constraint is studied in [29]. In this case, the design approach and the resulting precoding solution are similar to the HD case. The joint precoding and decoding design for an FD relay is studied in [17], [30], where both ZF and MMSE solutions are discussed. The ZF solution used in [17], [30] and most early works use a conventional approach based on the singular value decomposition (SVD) of the RSI channel. The main drawback of this approach is that the ZF solution only exists given that the numbers of antennas at the source, FD relay and the receiver satisfy a certain dimensionality condition. To overcome this limitation, [26] adopts an alternative criterion and proposes to maximize the signal-to-interference ratios between the power of the useful signal to the power of RSI at the relay input and output, respectively. Conventional ZF precoding and decoding are designed based on the singular vectors of the RSI channels. In [31], a joint design of ZF precoding and decoding is proposed to fully cancel the RSI at the relay, taking into account the source-relay and relay-destination channels. In [8] and [32], the precoding and decoding vectors are jointly optimized to maximize the end-to-end performance.

Buffer-aided schemes for decode-and-forward (DF)-FD Gaussian relay channels were proposed in [14], [33], [34]. In [33], the authors assumed that the RSI at the FD relay is negligible which is not realistic. The authors in [34] assumed that the RSI is fixed and does not vary with time. This may or may not be the case depending on system parameters and the employed self-interference cancellation techniques [35], [36]. In addition, the authors in [34] do not investigate the case when both the source and the relay transmit with a fixed rate in all time slots; a scenario which is investigated in this paper. The authors of [12], [14] proposed a hybrid HD/FD scheme to maximize the throughput of a relaying system for fixed-rate data transmissions. However, the authors neglected the fact that the relay knows its transmitted data signal and can do better in mitigating its impact as will be fully investigated in this paper.

Most of the aforementioned research assumed that the RSI is known but the data symbols are unknown. The first assumption is impractical since, by definition, the RSI is the remaining interference after applying all kinds of practically feasible interference mitigation techniques. Plausibility of the assumption that the data symbols are unknown depends on the operating scenario. For instance, in DF-FD relaying, the relay needs to know the entire codeword to know the transmitted sequence. Hence, it makes sense to assume that the symbols are unknown until the entire codeword is decoded. But if we assume that the relay has a buffer to store the data received from the source node, the relay will have its own data which are possibly different from the data that are currently received from the source. Hence, an FD mode can be applied and the entire transmitted sequence is known a priori by the relay.

In this paper, we consider a buffer-aided MIMO-FD Gaussian relay channel. Since the relay has a buffer, it knows the codewords that it transmits. Given this information, the contributions of this paper are summarized as follows

- We derive closed-form expressions for the achievable rates of the source-relay and relay-destination links when the RSI is changing slowly or quickly.
- We show that the buffer can help in completely canceling the impact of RSI for the case of slow RSI, when the buffer is non-empty. The maximum achievable rate of the source-relay link under the FD mode is that of the source-relay link without interference.
- For fast RSI, we show that the achievable rate of the source-relay link is degraded due to RSI and the degradation is quantified analytically. When the optimal precoder that maximizes the achievable rate of the relay-destination link is used, we derive a closed-form expression for the achievable rate of the source-relay link.

*Notation:* Unless otherwise stated, lower- and upper-case bold letters denote vectors and matrices, respectively. $\mathbf{I}_N$ denotes the identity matrix whose size is $N \times N$. $\mathbb{C}^{M \times N}$ denotes the set of all complex matrices of size $M \times N$. $\mathbf{0}_{M \times N}$ denotes the all-zero matrix with size $M \times N$. $(\cdot)^\top$, $(\cdot)^*$, and $(\cdot)^H$ denote transpose, complex conjugate, and Hermitian (i.e., complex-conjugate transpose) operations, respectively. $|\cdot|$ denotes the sum of the value in brackets. $\mathcal{CN}(x, y)$ denotes a complex circularly-symmetric Gaussian random variable with mean $x$ and variance $y$. $\mathbb{E}\{\cdot\}$ denotes statistical expectation. $\otimes$ is the Kronecker product. $\mathrm{diag} = \{\cdot\}$ denotes a diagonal matrix with the enclosed elements as its diagonal elements. $\mathrm{Trace}\{\cdot\}$ denotes the sum of the diagonal entries of the matrix enclosed in braces. $\mathrm{vec}\{\cdot\}$ converts the input $M \times N$ matrix into a column vector of size $MN \times 1$.

## II. SYSTEM MODEL AND MAIN ASSUMPTIONS

We consider a dual-hop DF-FD MIMO Gaussian relay channel, where a multi-antenna source node communicates with its multi-antenna destination node through an FD multi-antenna relay node, as shown in Fig. 1. Each node is equipped with $M$ antennas.[1] A direct link between the source and its destination (i.e., source-destination link) does not exist due to shadowing and large distances between them [14], [33], [34]. We assume that the relay node is equipped with a finite-size buffer/queue to store the incoming data traffic from the source node. We denote the buffer at the relay node as $Q_R$ and its maximum size as $Q_{max}$. The source node is always backlogged with data to transmit. It is assumed that the time is partitioned into discrete equal-size time slots of $T$ seconds, where the duration of one time slot is equal to the channel coherence time and the channel bandwidth is $W$. We use subscripts $S$, $R$, and $D$ to denote the source node, relay node, and destination node, respectively.

Each wireless link exhibits a quasi-static fading where a channel matrix between two nodes remains unchanged within the duration of one time slot and changes independently from one time slot to another. We consider slow-RSI and fast-RSI scenarios, where the RSI is fixed over the entire codeword or changes from one symbol duration to another within the codeword, respectively.[2] We denote the RSI coefficient matrix by $\mathbf{H}_{RR}$ in the slow-RSI case and by $\bar{\mathbf{H}}_{RR}$ in the fast-RSI case. The elements of the RSI coefficient matrices are independent and identically distributed (i.i.d.) zero-mean circularly-symmetric Gaussian-distributed random variables with variance $\sigma_{RR}^2$ [35]–[38]. Each link is also corrupted by an additive white Gaussian noise (AWGN) process with zero mean and variance $\kappa_m$. It is assumed that the average transmit power at Node $m \in \{S, R\}$ is $P_m$. For notation simplicity, unless otherwise stated, we drop the time slot (coherence time) index from the equations and use only the symbol duration index. However, it is worth mentioning that the channels between the source and the relay and between the relay and the destination are constant within each coherence time and changes independently from one coherence time to another.

---

[1]For simplicity of presentation, we assume equal number of antennas at all nodes. However, the same analysis can be easily extended to the scenario of different number of antennas at all nodes.

[2]Typically, slow-RSI is assumed in the literature (e.g. [14], [33]), which represents an optimistic assumption and the best-case scenario in system design. However, in this paper, we investigate both scenarios of slow-/fast-RSI.

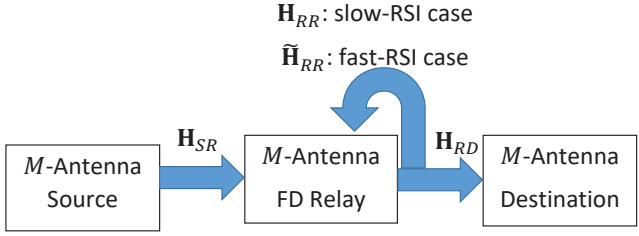

Figure 1. The considered dual-hop network. In the figure, we denote the RSI coefficient matrix by $\mathbf{H}_{RR} \in \mathbb{C}^{M \times M}$ in the slow-RSI case and by $\tilde{\mathbf{H}}_{RR} \in \mathbb{C}^{nM \times nM}$ in the fast-RSI case. The two matrices are different, since in the slow-RSI case, the RSI matrix remains constant over the entire codeword, while in the fast-RSI case, the RSI matrix changes from one symbol duration to another within the codeword. $\mathbf{H}_{SR} \in \mathbb{C}^{M \times M}$ and $\mathbf{H}_{RD} \in \mathbb{C}^{M \times M}$ are the channel matrices of the source-relay link and the relay-destination link, respectively.

The RSI channel is time-varying even when the communication links do not exhibit fading [17], [37], [39]–[41]. The RSI variations are due to the cumulative effects of various distortion sources including noise, carrier frequency offset, oscillator phase noise, analog-to-digital/digital-to-analog conversion (ADC/DAC) imperfections, in-phase/quadrature (I/Q) imbalance, imperfect channel estimation, etc [17], [37], [39]–[41]. These impairments and distortions have a significant impact on the RSI channel due to the very small distance between the transmitter-end and the receiver-end of the LSI channel. Moreover, the variations of the RSI channel are random and thereby cannot be accurately estimated at the FD node [17], [37], [39]–[41]. The statistical properties of the RSI variations are dependent on the hardware configuration and the adopted LSI suppression techniques. In [37], the RSI is assumed to be fixed/constant during the transmission of a codeword comprised of many symbols. Hence, the RSI model proposed in [37], and most of the papers in the literature, captures only the long-term, i.e., codeword-by-codeword, statistical properties of the RSI channel. However, the symbol-by-symbol RSI variations are not captured by the model proposed in [37] since these variations are averaged out. Nevertheless, for a meaningful information-theoretical analysis, the symbol-by-symbol variations of the RSI should be taken into consideration. The statistics of the RSI variations affect the achievable rates of the considered FD Gaussian relay channels. In this paper, we derive the achievable rates of the considered Gaussian relay channels for both the best-case RSI model (slow-RSI case) and the worst-case RSI model (fast-

RSI case). In addition, the slow-RSI model is suitable for the cases of fixed-rate transmission and when analyzing the system based on average performance [41]. Hence, it will be adopted when we study the fixed-rate transmission scenario in Section V.

In the following sections, we derive the closed-form expressions for the achievable rate of the source-relay link for slow-RSI and fast-RSI cases.

## III. Slow-RSI Case

In this case, the RSI varies across time slots (i.e., from one coherence time duration to another), but remains fixed within each time slot.

### A. Achievable Rates Derivations

The achievable rates of the communication links in buffer-aided relay networks change based on the relaying queue state (i.e., empty or non-empty). That is, if the relay's queue is empty, FD mode operation is not possible since, as mentioned in [42], the practicality of DF-FD relaying is questionable, when the relay does not have the entire codeword prior to data transmission. Hence, we simply assume that, when the relay is empty, it operates in an HD mode and it receives data of the source node. Assume that the source node transmits $M$ independent codewords of length $n$, $n > M$. The data matrix transmitted by the source node, denoted by $\mathbf{X}_S \in \mathbb{C}^{n \times M}$, is given by

$$\mathbf{X}_S = \begin{pmatrix} X_{S,1}(1) & X_{S,2}(1) & \dots & X_{S,M}(1) \\ X_{S,1}(2) & X_{S,2}(2) & \dots & X_{S,M}(2) \\ \vdots & \vdots & \vdots & \vdots \\ X_{S,1}(n) & X_{S,2}(n) & \dots & X_{S,M}(n) \end{pmatrix}, \quad (1)$$

where the elements of $\mathbf{X}_S$ are assumed to be i.i.d. Gaussian circularly-symmetric random variables with zero mean and variance $\tilde{P}_S = P_S/M$ (i.e., variance per data stream). We assume Gaussian data signals at the source and relay nodes as in, e.g., [14], [33], [35]–[38] and the references therein. The received signal at the relay is given by

$$\mathbf{Y}_R = \mathbf{X}_S \mathbf{H}_{SR}^\top + \boldsymbol{\epsilon}_R, \quad (2)$$

where $\boldsymbol{\epsilon}_R \in \mathbb{C}^{n \times M}$ is the AWGN noise matrix at the relay, and $\mathbf{H}_{SR} \in \mathbb{C}^{M \times M}$ is the channel matrix between the source node and the relay node with element $(v, \ell)$ in $\mathbf{H}_{SR}$ being the channel coefficient between the source's $v$-th antenna and the relay's $\ell$-th antenna. Hence, the achievable rate of the source-relay link is given by

$$\mathcal{I}_{SR}^{\text{HD}} = \sum_{v=1}^{M} \log_2 \left( 1 + \frac{\tilde{P}_S}{\kappa_R} \eta_v \right), \quad (3)$$

where $\eta_v$ is the $v$-th eigenvalue of $\mathbf{H}_{SR} \mathbf{H}_{SR}^H$. The expression in (3) can be deduced from Appendix A by setting the relay's data precoding matrix to zero (i.e., $\boldsymbol{\Psi} = \mathbf{0}_{M \times M}$).

If the relay's queue is non-empty, the relay transmits a codeword that is different from the source's codeword and operation in the FD mode is possible. The received signal at the relay's receiver is given by

$$\mathbf{Y}_R = \mathbf{X}_S \mathbf{H}_{SR}^\top + \mathbf{X}_R \boldsymbol{\Psi} \mathbf{H}_{RR}^\top + \boldsymbol{\epsilon}_R, \tag{4}$$

where $\mathbf{X}_R \in \mathbb{C}^{n \times M}$ is the data matrix transmitted by the relay node and has the same structure as $\mathbf{X}_S$ in (1) but with the codewords transmitted by the relay which are independent from those transmitted by the source node. The elements of $\mathbf{X}_R$ are i.i.d. with $\frac{1}{n}\mathbb{E}\{\text{vec}\{\mathbf{X}_R\}^H \text{vec}\{\mathbf{X}_R\}\} = P_R$ and, hence, a realization of $\mathbf{X}_R$ has a rank equal to $M$ with probability one. The relay node applies the data precoding matrix $\boldsymbol{\Psi} \in \mathbb{C}^{M \times M}$ to its data vectors, where $\text{Trace}\{\boldsymbol{\Psi}\boldsymbol{\Psi}^H\} = M$. Matrix $\mathbf{H}_{RR} \in \mathbb{C}^{M \times M}$ is the RSI coefficient matrix. Element $(v, \ell)$ in $\mathbf{H}_{RR}^\top$ represents how transmission from the relay's $\ell$-th antenna impacts the relay's received signal on its $v$-th antenna. We collect the elements of the matrix $\mathbf{Y}_R$ in a vector to compute the information rate under Gaussian signaling as follows [43]

$$\mathcal{I}_{SR}^{\text{FD}} = \frac{1}{n} \Big[ \log_2 \det \big( \mathbb{E}\left\{ \text{vec}\left\{ \mathbf{Y}_R \right\} \text{vec}\left\{ \mathbf{Y}_R \right\}^H \right\} \big)$$
$$- \log_2 \det \big( \mathbb{E}\left\{ \text{vec}\left\{ \mathbf{X}_R \boldsymbol{\Psi} \mathbf{H}_{RR}^\top + \boldsymbol{\epsilon}_R \right\} \right. \tag{5}$$
$$\left. \times \left( \text{vec}\left\{ \mathbf{X}_R \boldsymbol{\Psi} \mathbf{H}_{RR}^\top + \boldsymbol{\epsilon}_R \right\} \right)^H \right\} \big) \Big],$$

where

$$\text{vec}\{\mathbf{Y}_R\} = (\mathbf{H}_{SR} \otimes \mathbf{I}_n)\,\text{vec}\{\mathbf{X}_S\} + (\mathbf{I}_M \otimes \mathbf{X}_R \boldsymbol{\Psi})\,\text{vec}\{\mathbf{H}_{RR}^\top\}$$
$$+ \text{vec}\{\boldsymbol{\epsilon}_R\}, \tag{6}$$

by exploiting the following property of $\text{vec}\{.\}$

$$\text{vec}\{\mathbf{AB}\} = \left( \mathbf{B}^\top \otimes \mathbf{I}_n \right) \text{vec}\{\mathbf{A}\} = (\mathbf{I}_M \otimes \mathbf{A})\,\text{vec}\{\mathbf{B}\}, \tag{7}$$

where $\mathbf{A} \in \mathbb{C}^{n \times l}$ and $\mathbf{B} \in \mathbb{C}^{l \times M}$.

**Proposition 1.** *The information rate of the source-relay link under Gaussian signaling is given by*

$$\mathcal{I}_{SR}^{\text{FD}} = \frac{1}{n} \sum_{v=1}^{M} \Big( \log_2 \det \big( \tilde{P}_S \eta_v \mathbf{I}_n + \sigma_{RR}^2 \mathbf{X}_R \boldsymbol{\Psi} \boldsymbol{\Psi}^H \mathbf{X}_R^H + \kappa_R \mathbf{I}_n \big)$$
$$- \log_2 \det \big( \sigma_{RR}^2 \mathbf{X}_R \boldsymbol{\Psi} \boldsymbol{\Psi}^H \mathbf{X}_R^H + \kappa_R \mathbf{I}_n \big) \Big), \tag{8}$$

*where $\eta_v$ is the $v$-th eigenvalue of $\mathbf{H}_{SR} \mathbf{H}_{SR}^H$.*

*Proof.* See Appendix A. □

**Proposition 2.** *The precoder that maximizes the information rate of the source-relay link under Gaussian relay channel, which is referred to as the $\mathcal{I}_{SR}^{\text{FD}}$-maximizing precoder, is rank-1.*

*Proof.* See Appendix B. □

In the next two propositions, we present a closed-form expression for the achievable rate of the source-relay link when the relay uses the rank-1 precoder (i.e., $\mathcal{I}_{SR}^{\text{FD}}$-maximizing precoder) in Proposition 2 and the $\mathcal{I}_{RD}^{\text{FD}}$-maximizing precoder derived in Appendix C, respectively.

**Proposition 3.** *Letting $\boldsymbol{\Psi} = \sqrt{M}\mathbf{q}\mathbf{q}^H$, where $\mathbf{q} \in \mathbb{C}^{M \times 1}$ with $\mathbf{q}^H \mathbf{q} = 1$, and substituting with $\boldsymbol{\Psi} = \sqrt{M}\mathbf{q}\mathbf{q}^H$ into (8), the*

*information rate of the source-relay link under the slow-RSI scenario is given by*

$$\mathcal{I}_{SR}^{\text{FD}} = \sum_{v=1}^{M} \log_2 \left( 1 + \frac{\tilde{P}_S}{\kappa_R} \eta_v \right)$$
$$+ \frac{1}{n} \sum_{v=1}^{M} \Big( \log_2 \left( 1 + \frac{\sigma_{RR}^2}{(\tilde{P}_S \eta_v + \kappa_R)} \mathbf{q}^H \mathbf{X}_R^H \mathbf{X}_R \mathbf{q} \right) \tag{9}$$
$$- \log_2 \left( 1 + \frac{\sigma_{RR}^2}{\kappa_R} \mathbf{q}^H \mathbf{X}_R^H \mathbf{X}_R \mathbf{q} \right) \Big),$$

*where $\mathbf{q}$ is the normalized eigenvector corresponding to the minimum eigenvalue of $\mathbf{X}_R^H \mathbf{X}_R$.*

*Proof.* See Appendix D. □

**Proposition 4.** *When the optimal precoder that maximizes the information rate of the relay-destination channel derived in Appendix C is used by the relay node, the achievable rate expression of the source-relay link can be rewritten as*

$$\mathcal{I}_{SR}^{\text{FD}} = \frac{1}{n} \sum_{v=1}^{M} \Big( n \log_2(1 + \tilde{P}_S \eta_v)$$
$$+ \log_2 \det \left( \mathbf{I}_M + \frac{\sigma_{RR}^2}{\tilde{P}_S \eta_v + \kappa_R} \mathbf{E} \mathbf{E}^H \mathbf{X}_R^H \mathbf{X}_R \right) \tag{10}$$
$$- \log_2 \det \left( \mathbf{I}_M + \frac{\sigma_{RR}^2}{\kappa_R} \mathbf{E} \mathbf{E}^H \mathbf{X}_R^H \mathbf{X}_R \right) \Big),$$

*where $\boldsymbol{\Psi} = \mathbf{E} \mathbf{Q}_{RD}^H$ is full rank with $\mathbf{E}$ denoting a diagonal matrix such that $\mathbf{E}\mathbf{E}^H$ contains the power fractions assigned to each data stream and $\text{Trace}\{\boldsymbol{\Psi}\boldsymbol{\Psi}^H\} = M$.*

*Proof.* See Appendix E. □

**Proposition 5.** *When $n$ goes to infinity, the achievable rate in the slow-RSI case is given by*

$$\mathcal{I}_{SR}^{\text{FD}} = \sum_{v=1}^{M} \log_2 \left( 1 + \frac{\tilde{P}_S}{\kappa_R} \eta_v \right), \tag{11}$$

*which is the achievable rate of the source-relay channel with no interference.*

*Proof.* When $n \to \infty$, the diagonal elements of $\frac{1}{n}\mathbf{X}_R^H \mathbf{X}_R$ in (10) converge to $P_R$ and the off-diagonal elements scaled by $1/n$ converge to zero almost surely [44]. Thus,

$$\mathcal{I}_{SR}^{\text{FD}} = \lim_{n \to \infty} \frac{1}{n} \sum_{v=1}^{M} \Big( n \log_2 \left( 1 + \frac{\tilde{P}_S}{\kappa_R} \eta_v \right)$$
$$+ \sum_{\ell=1}^{M} \log_2 \left( 1 + \frac{\sigma_{RR}^2}{\tilde{P}_S \eta_v + \kappa_R} n P_R |E_\ell|^2 \right) \tag{12}$$
$$- \sum_{\ell=1}^{M} \log_2 \left( 1 + \frac{\sigma_{RR}^2}{\kappa_R} n P_R |E_\ell|^2 \right) \Big),$$

where $E_\ell$ is the $\ell$-th element on the main diagonal of $\mathbf{E}$. The last two terms go to zero for finite $M$ ($n \gg M$). Thus, we get the expression in (11). □

The result in Proposition 5 is promising since it implies that, regardless of the precoder employed at the relay, the achievable rate of the source-relay channel under FD operation equals to the rate of the source-relay channel with no interference. Hence, if the relay uses the precoder that maximizes the achievable rate of the relay-destination link (i.e., $\mathcal{I}_{RD}^{\mathrm{FD}}$-maximizing precoder derived in Appendix C), the achievable rates of the two links (i.e., source-relay and relay-destination links) will be simultaneously maximized. Accordingly, the channel capacities of the two links (source-relay and relay-destination links) can be achieved.

## IV. FAST-RSI CASE

In the case of fast-RSI, the RSI changes independently from one symbol duration to another. That is, each symbol within the codeword experiences a different RSI realization.

### A. Achievable Rates Derivations

Assuming $M$ independent codewords transmitted by the relay node, the data matrix, denoted by $\tilde{\mathbf{X}}_R \in \mathbb{C}^{n \times nM}$, is given by

$$\tilde{\mathbf{X}}_R = \begin{pmatrix} \mathbf{X}_R(1) & \mathbf{0} & \dots & \mathbf{0} \\ \mathbf{0} & \mathbf{X}_R(2) & \dots & \mathbf{0} \\ \vdots & \vdots & \ddots & \vdots \\ \mathbf{0} & \mathbf{0} & \dots & \mathbf{X}_R(n) \end{pmatrix}, \quad (13)$$

where $\mathbf{X}_R(j) = [X_{R,1}(j)\ X_{R,2}(j)\ \dots\ X_{R,M}(j)] \in \mathbb{C}^{1 \times M}$ is the data symbols vector transmitted by the relay at the $j$th symbol duration. The RSI coefficient matrix, denoted by $\tilde{\mathbf{H}}_{RR} \in \mathbb{C}^{nM \times M}$, is given by

$$\tilde{\mathbf{H}}_{RR} = \begin{pmatrix} \tilde{\mathbf{H}}_{RR(1)}^{\top} \\ \tilde{\mathbf{H}}_{RR(2)}^{\top} \\ \vdots \\ \tilde{\mathbf{H}}_{RR(n)}^{\top} \end{pmatrix}, \quad (14)$$

where each block $\tilde{\mathbf{H}}_{RR}(j)$ is $M \times M$.

Hence, the received signal vector at the relay's receiver

$$\mathbf{Y}_R = \mathbf{X}_S \mathbf{H}_{SR}^{\top} + \tilde{\mathbf{X}}_R \tilde{\boldsymbol{\Psi}}_R \tilde{\mathbf{H}}_{RR} + \boldsymbol{\epsilon}_R, \quad (15)$$

where $\boldsymbol{\epsilon}_R \in \mathbb{C}^{N \times M}$ is the noise matrix at the relay and $\tilde{\boldsymbol{\Psi}}_R \in \mathbb{C}^{nM \times nM}$ is the data precoding matrix used at the relay node, $\tilde{\mathbf{H}}_{RR} \in \mathbb{C}^{nM \times nM}$ is the RSI channel matrix. Matrix $\tilde{\boldsymbol{\Psi}}_R \in \mathbb{C}^{nM \times nM}$ has the block diagonal structure $\tilde{\boldsymbol{\Psi}}_R = \mathrm{diag}\{\boldsymbol{\Phi}, \boldsymbol{\Phi}, \dots, \boldsymbol{\Phi}\}$, where $\boldsymbol{\Phi}$ is an $M \times M$ matrix with $\mathrm{Trace}\{\boldsymbol{\Phi}\boldsymbol{\Phi}^H\} = M$.

**Proposition 6.** *The achievable rate of the source-relay link for the fast-RSI case is given by*

$$\mathcal{I}_{SR}^{\mathrm{FD}} = \frac{1}{n} \sum_{v=1}^{M} \Big( \log_2 \det \Big( \tilde{P}_S \eta_v \mathbf{I}_n + \sigma_{RR}^2 \tilde{\mathbf{X}}_R \tilde{\boldsymbol{\Psi}} \left( \tilde{\mathbf{X}}_R \tilde{\boldsymbol{\Psi}} \right)^H + \kappa_R \mathbf{I}_n \Big)$$
$$- \log_2 \det \Big( \sigma_{RR}^2 \tilde{\mathbf{X}}_R \tilde{\boldsymbol{\Psi}} \left( \tilde{\mathbf{X}}_R \tilde{\boldsymbol{\Psi}} \right)^H + \kappa_R \mathbf{I}_n \Big) \Big). \quad (16)$$

*Proof.* See Appendix F. □

Using the same approach as in the slow-RSI case to derive the precoder that maximizes the information rate of source-relay link (i.e., $\mathcal{I}_{SR}^{\mathrm{FD}}$-maximizing precoder), if $\tilde{\boldsymbol{\Psi}}_R = \mathrm{diag}\{\boldsymbol{\Phi}_1, \boldsymbol{\Phi}_2, \dots, \boldsymbol{\Phi}_N\}$, then choosing $\boldsymbol{\Phi}_k$ to be $\left( \mathbf{I}_M - \frac{\mathbf{X}_R^*(j)\mathbf{X}_R^{\top}(j)}{\|\mathbf{X}_R(j)\|^2} \right)$ nulls the relay's transmission. In other words, although this precoder cancels the relay's transmission from the relay's receive side, but it also cancels the transmissions from everywhere else. Thus, this precoder reduces the achievable rate of the relay-destination link to zero, and effectively makes the relay operate as an *HD terminal*. In the sequel, we study the achievable rates of the source-relay and relay-destination links, respectively, when the precoder $\boldsymbol{\Phi}$ follows the $\mathcal{I}_{SR}^{\mathrm{FD}}$-maximizing and $\mathcal{I}_{RD}^{\mathrm{FD}}$-maximizing precoder designs employed in the slow-RSI case. Moreover, we study the asymptotic case as $n \to \infty$.

Note that the $\mathcal{I}_{SR}^{\mathrm{FD}}$-maximizing precoder is not necessarily the precoder that also maximizes the achievable rate of the relay-destination link. The optimal precoder at the relay should be designed based on a selected performance criterion (e.g., maximum rate between the two rates of the communications hops, minimum end-to-end bit-error-rate probability, maximum sum-rate of the two communication hops, etc). For example, if the goal is to maximize the minimum between the information rates of the two hops (i.e. maximize $\min\{\mathcal{I}_{SR}^{\mathrm{FD}}, \mathcal{I}_{RD}^{\mathrm{FD}}\}$), we need to derive the optimal precoder based on that. That is, we need to find the optimal precoder that maximizes $\mathcal{I}_{RD}^{\mathrm{FD}}$. However, this precoder is difficult to obtain analytically even for $M = 2$. To gain some insights, we provide a heuristic solution which is realized as follows. The relay uses the two precoders: $\mathcal{I}_{SR}^{\mathrm{FD}}$-maximizing and $\mathcal{I}_{RD}^{\mathrm{FD}}$-maximizing precoders. Then, it computes the minimum achievable rates of the two hops under each case. After that, the relay selects the precoder with the highest minimum achievable rate.

**Proposition 7.** *The achievable rate of the source-relay link for the fast-RSI case, when the relay uses the $\mathcal{I}_{SR}^{\mathrm{FD}}$-maximizing precoder of the slow-RSI case, which has the form $\boldsymbol{\Phi} = \sqrt{M}\mathbf{q}\mathbf{q}^H$, is given by*

$$\mathcal{I}_{SR}^{\mathrm{FD}} = \sum_{v=1}^{M} \log_2 \left( 1 + \frac{\tilde{P}_S}{\kappa_R} \eta_v \right) + \sum_{v=1}^{M} \mathbb{E}\left\{ \log_2 \left( 1 + \frac{\frac{\sigma_{RR}^2}{\kappa_R}}{1 + \frac{\tilde{P}_S}{\kappa_R}\eta_v} |\mathbf{X}_R(j)\mathbf{q}|^2 \right) - \log_2 \left( 1 + \frac{\sigma_{RR}^2}{\kappa_R} |\mathbf{X}_R(j)\mathbf{q}|^2 \right) \right\}. \quad (17)$$

*Proof.* See Appendix G. □

The next proposition considers the case where the relay uses the precoder that maximizes the achievable rate of the relay-destination link (i.e., the $\mathcal{I}_{RD}^{\mathrm{FD}}$-maximizing precoder).

**Proposition 8.** *The achievable rate of the source-relay link for the fast-RSI case, when the relay uses the $\mathcal{I}_{RD}^{\mathrm{FD}}$-maximizing precoder of the slow-RSI, which has the form $\boldsymbol{\Phi} = \mathbf{E}\mathbf{Q}_{RD}^*$, is given by*

$$\mathcal{I}_{SR}^{\mathrm{FD}} = \sum_{v=1}^{M} \log_2 \left( 1 + \frac{\tilde{P}_S}{\kappa_R} \eta_v \right)$$
$$+ \sum_{v=1}^{M} \left( \mathbb{E}\left\{ \log_2 \left( 1 + \frac{\frac{\sigma_{RR}^2}{M\kappa_R}\sum_{i=1}^{M}|X_{R,i}(j)|^2}{1 + \frac{\tilde{P}_S}{\kappa_R}\eta_v} \right) \right\} \right.$$
$$\left. - \mathbb{E}\left\{ \log_2 \left( 1 + \frac{\sigma_{RR}^2}{M\kappa_R}\sum_{i=1}^{M}|X_{R,i}(j)|^2 \right) \right\} \right). \quad (18)$$

*Proof.* See Appendix H. □

For the case of equal power allocation to data streams, when $M$ is large, we can approximate $\mathcal{X}(j) \approx \sum_{i=1}^{M} |X_{R,i}(j)|^2 = MP_R$ from the strong law of large numbers. Hence, the achievable rate of the source-relay link is

$$\mathcal{I}_{SR}^{FD} = \sum_{v=1}^{M} \left( \log_2 \left( \frac{\tilde{P}_S}{\kappa_R} \eta_v + \frac{\sigma_{RR}^2 P_R}{\kappa_R} + 1 \right) - \log_2 \left( \frac{\sigma_{RR}^2 P_R}{\kappa_R} + 1 \right) \right). \tag{19}$$

**A special case- single-input single-output (SISO):**
Let $h_{SR}$ and $h_{RD}$ denote the channel coefficients of the source-relay and relay-destination links, respectively. Since $n$ is very large, from the strong law of large numbers, $\frac{1}{n} \sum_{i=1}^{n} \log_2 \left(1 + \gamma_q |x_R(i)|^2\right)$ will almost surely converge to $\mathbb{E} \left\{ \log_2 \left(1 + \gamma_q |x_R(i)|^2\right) \right\}$ where $\gamma_q \in \{\gamma_1, \gamma_2\}$ with $\gamma_1 = \frac{1}{\kappa_R + \tilde{P}_S |h_{SR}|^2}$ and $\gamma_2 = \frac{1}{\kappa_R}$. Since $|x_R(i)|^2$ is an exponentially-distributed random variable, the average of $\log_2 \left(1 + \gamma_q |x_R(i)|^2\right)$ is given by

$$\begin{aligned}
\mathbb{E} \left\{ \log_2 \left(1 + \gamma_q |x_R(i)|^2\right) \right\} &= \int_0^\infty \log_2 \left(1 + \gamma_q |x_R(i)|^2\right) d|x_R(i)|^2 \\
&= \frac{\exp\left(\frac{1}{\gamma_q \sigma_{RR}^2 P_R}\right)}{\ln(2)} \text{Ei}\left(\frac{1}{\gamma_q \sigma_{RR}^2 P_R}\right),
\end{aligned} \tag{20}$$

where $\text{Ei}(x) = \int_x^\infty \frac{\exp(-u)}{u} du$ is the exponential integral. Substituting in the information rate expression of the source-relay link, we have

$$\begin{aligned}
\mathcal{I}_{SR}^{FD} = {} & \log_2 \left(1 + |h_{SR}|^2 \frac{\tilde{P}_S}{\kappa_R}\right) \\
& + \frac{1}{\ln(2)} \exp\left(\frac{1}{\gamma_1 \sigma_{RR}^2 P_R}\right) \text{Ei}\left(\frac{1}{\gamma_1 \sigma_{RR}^2 P_R}\right) \\
& - \frac{1}{\ln(2)} \exp\left(\frac{1}{\gamma_2 \sigma_{RR}^2 P_R}\right) \text{Ei}\left(\frac{1}{\gamma_2 \sigma_{RR}^2 P_R}\right).
\end{aligned} \tag{21}$$

The achievable rate of the relay-destination link is given by

$$\mathcal{I}_{RD}^{FD} = \mathcal{I}_{RD}^{HD} = \log_2 \left(1 + |h_{RD}|^2 \frac{P_R}{\kappa_D}\right). \tag{22}$$

## V. A Case Study: Fixed-Rate Transmission

In this section, we study the fixed-rate transmission case where the source and relay transmit with a fixed rate of $\mathcal{R}$ bits/sec/Hz. Since we assume fixed-rate transmissions under queueing constraints, the RSI channel is assumed to be slow to capture only the long-term, i.e., codeword-by-codeword, statistical properties [41]. The relaying queue can be modeled as a birth-death process since only one packet is decoded at the relay, one packet is transmitted by the relay, or one packet is decoded and one packet is transmitted by the relay at the same time.

When the relaying queue is empty, the probability that the source packet is correctly decoded and stored at the relay (i.e., the queue state transits from state 0 to state 1) is given by

$$a_0 = \Pr\left\{\mathcal{I}_{SR}^{HD} \geq \mathcal{R}\right\}. \tag{23}$$

If the relaying queue is non-empty, the optimal transmission scheme is that both the source and the relay transmit data simultaneously. This is because the two links (i.e., source-relay and relay-destination links) are completely independent and separable because, when RSI is slow and $n \to \infty$, the self-interference at the relay is removed and the source-relay channel is not affected by relay transmissions. The probability that the queue transits from state $\ell > 0$ to state $\ell + 1$, denoted by $a_\ell$, is equal to the probability that the source-relay link is not in outage and that of the relay-destination link is in outage. Hence, $a_\ell$ is given by

$$a_\ell = a = \Pr\left\{\mathcal{I}_{SR}^{FD} \geq \mathcal{R}\right\} \Pr\left\{\mathcal{I}_{RD}^{FD} < \mathcal{R}\right\}, \tag{24}$$

where $\ell > 0$ and it denotes the state of the relaying queue (i.e., number of packets at the relaying queue) and $\mathcal{I}_{RD}^{FD}$ is the achievable rate of the source-destination link which is derived in Appendix C. Similarly, the probability that the queue transits from state $0 < \ell < Q_{\max}$ to state $\ell - 1$, denoted by $b_\ell$, is equal to the probability that the source-relay link is in outage, whereas the relay-destination link is not. Hence, $b_\ell$ is given by

$$b_\ell = b = \Pr\left\{\mathcal{I}_{RD}^{FD} \geq \mathcal{R}\right\} \Pr\left\{\mathcal{I}_{SR}^{FD} < \mathcal{R}\right\}. \tag{25}$$

When the relaying buffer is full, the transition probability, denoted by $b_{Q_{\max}}$, is given by

$$b_{Q_{\max}} = \Pr\left\{\mathcal{I}_{RD}^{FD} \geq \mathcal{R}\right\}, \tag{26}$$

since the relay cannot accept any new packets before delivering the ones stored in its buffer.

Analyzing the relaying queue Markov chain as in [4], the local balance equations are given by

$$\beta_\nu a_\nu = \beta_{\nu+1} b_{\nu+1}, 0 \leq \nu \leq Q_{\max} - 1, \tag{27}$$

where $\beta_\nu$ denotes the probability of having $\nu$ packets in the relaying queue. Using the balance equations recursively, the stationary distribution of $\beta_\nu$ is given by

$$\beta_\nu = \beta_0 \prod_{\varrho=0}^{\nu-1} \frac{a_\varrho}{b_{\varrho+1}}, \tag{28}$$

where $\beta_0 = \left(1 + \sum_{\nu=1}^{Q_{\max}} \prod_{\varrho=0}^{\nu-1} \frac{a_\varrho}{b_{\varrho+1}}\right)^{-1}$ is obtained using the normalization condition $\sum_{\nu=0}^{Q_{\max}} \beta_\nu = 1$.

By using the normalization condition, we get

$$\beta_\nu = \beta_0 \prod_{\varrho=0}^{\nu-1} \frac{a_\varrho}{b_{\varrho+1}} = \begin{cases} \beta_0 \frac{a_0}{a} \frac{a^\nu}{b^\nu}, & \nu < Q_{\max} \\ \beta_0 \frac{a_0 b}{ab_{Q_{\max}}} \frac{a^{Q_{\max}}}{b^{Q_{\max}}}, & \nu = Q_{\max} \end{cases}. \tag{29}$$

The probability of the queue being empty is given by

$$\begin{aligned}
\beta_0 &= \left(1 + \frac{a_0}{a} \left(\sum_{\nu=1}^{Q_{\max}-1} \left(\frac{a}{b}\right)^\nu + \frac{b}{b_{Q_{\max}}} \left(\frac{a}{b}\right)^{Q_{\max}}\right)\right)^{-1} \\
&= \left(1 + \frac{a_0}{b} \frac{1 - \left(\frac{a}{b}\right)^{Q_{\max}-1}}{1 - \left(\frac{a}{b}\right)} + \frac{a_0 b}{ab_{Q_{\max}}} \left(\frac{a}{b}\right)^{Q_{\max}}\right)^{-1}.
\end{aligned} \tag{30}$$

If the queue is unlimited in size (i.e., $Q_{\max} \to \infty$), $a < b$ is a necessary condition for the queue stability and for the steady-state solution to exist. Simplifying Eqn. (30), we get

$$\beta_0 = \left(1 + \frac{a_0}{b} \frac{1}{1 - \left(\frac{a}{b}\right)}\right)^{-1} = \frac{b - a}{b - a + a_0}. \tag{31}$$

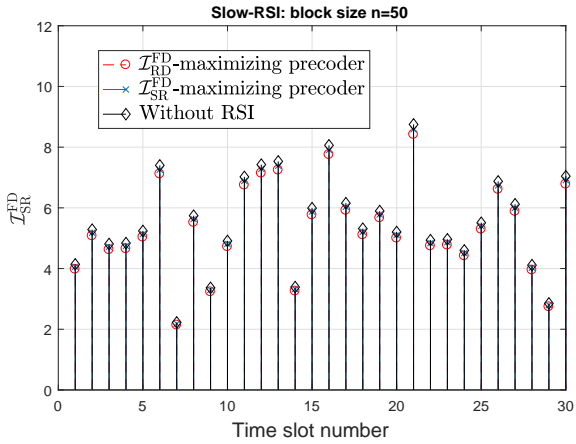

Figure 2. Achievable rate of the source-relay channel for the slow-RSI scenario when block size is $n = 50$.

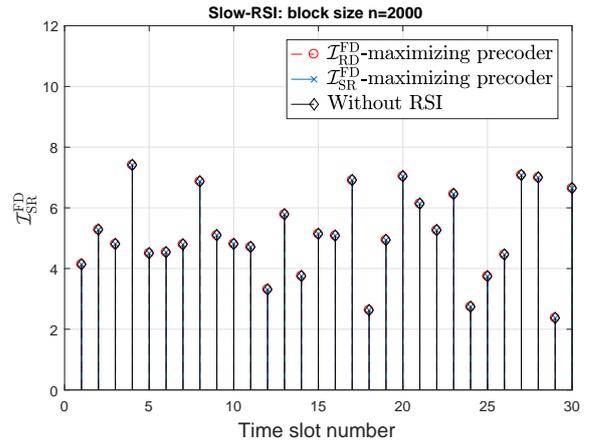

Figure 3. Achievable rate of the source-relay channel for the slow-RSI scenario when block size is $n = 2000$.

The system throughput in packets/slot, which is the number of correctly decoded packets at the destination per time slot, is given by

$$\mu_d = (1 - \beta_0) \Pr\{\mathcal{I}_{RD}^{FD} \geq \mathcal{R}\}, \tag{32}$$

which represents the probability that the queue is non-empty and that the relay-destination link is not in outage.

## VI. NUMERICAL RESULTS AND SIMULATIONS

In this section, we verify the analytical findings in each of the investigated scenarios. We start with the slow-RSI case followed by the fast-RSI case. Then, we show numerical results for the case of fixed-rate transmission. Unless otherwise stated, we use the following system's parameters to generate the results: the fading channels are assumed to be complex circularly-symmetric Gaussian random variables with zero mean and unit variance, $\kappa_R = \kappa_D = \kappa$, $P_S/\kappa = 10$ dB, $P_R/\kappa = 10$ dB, and $\sigma_{RR}^2 = 0$ dB.

### A. Slow-RSI Case

To verify our derivations, we provide some numerical results for the achievable rate in the case of slow RSI. Our main message from the numerical results in this subsection is to verify that the optimal precoder that maximizes the achievable rate of the source-relay link in case of finite block size $n$ is the rank-1 precoder (which we refer to as the $\mathcal{I}_{RD}^{FD}$-maximizing precoder). Moreover, we want to verify that when the block size is sufficiently large, any precoder can be used, including the one that maximizes the achievable rate of the relay-destination link, with no rate loss (i.e., the information rate under slow RSI converges to the information rate of the no interference case). Figs. 2 and 3 show the achievable rate of the source-relay link for both cases of $\mathcal{I}_{SR}^{FD}$-maximizing and $\mathcal{I}_{RD}^{FD}$-maximizing precoders when the block size is finite and equal to $n = 50$ and $n = 2000$ symbols, respectively. We also show the maximum achievable rate for the source-relay link when the RSI is zero. Figs. 2 and 3 are generated using unit-variance channels, $M = 2$, and the instantaneous randomly-generated channel matrices in Table I for three time slots.

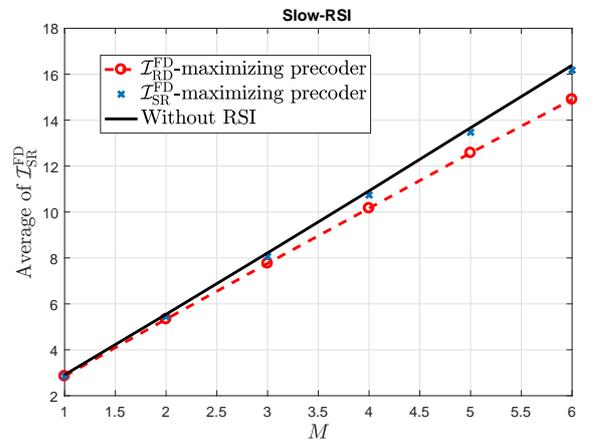

Figure 4. Average achievable rate of source-relay link for the slow-RSI scenario.

As shown in Fig. 2, the $\mathcal{I}_{SR}^{FD}$-maximizing precoder achieves a rate which is closer to the upper bound than that of the $\mathcal{I}_{RD}^{FD}$-maximizing precoder. In Fig. 3, all curves overlap thereby implying that for slow-RSI regardless of the used precoder at the relay, the RSI is completely canceled when $n$ is sufficiently high; which verifies our theoretical findings. For the case of $M > 2$ and due to the significant increase in the number of system's parameters and channel matrices, we plot the average achievable rate versus $M$ in Fig. 4. As it can be seen from the figure, the $\mathcal{I}_{SR}^{FD}$-maximizing precoder achieves almost the no-interference achievable rate when the block size is finite, i.e., $n = 50$. Increasing the number of antennas increases the achievable rate of the source-relay link.

### B. Fast-RSI Case

We evaluate the achievable rate expressions that we obtained for the fast-RSI scenario. First, we present some numerical results for the instantaneous achievable rate expressions by using Table I for the case of $M = 2$. Then, for the case of $M > 2$ and since the size of the channel matrices and the system's parameters increase significantly, we present the average of the achievable rate expressions, averaged across channel realizations, versus $M$. In Fig. 5, we show the achievable rate of the fast-RSI scenario when both the $\mathcal{I}_{\mathrm{SR}}^{\mathrm{FD}}$-maximizing and $\mathcal{I}_{\mathrm{RD}}^{\mathrm{FD}}$-maximizing precoders of the slow-RSI are used by the relay. As expected, the $\mathcal{I}_{\mathrm{SR}}^{\mathrm{FD}}$-maximizing precoder achieves a higher source-relay link achievable rate than the $\mathcal{I}_{\mathrm{RD}}^{\mathrm{FD}}$-maximizing precoder. This is because the $\mathcal{I}_{\mathrm{SR}}^{\mathrm{FD}}$-maximizing precoder decreases the interference caused by the data transmissions at the relay. Fig. 6 shows the average achievable rate of the source-relay link versus $M$ for the cases of $\mathcal{I}_{\mathrm{RD}}^{\mathrm{FD}}$-maximizing and $\mathcal{I}_{\mathrm{SR}}^{\mathrm{FD}}$-maximizing precoders. The $\mathcal{I}_{\mathrm{SR}}^{\mathrm{FD}}$-maximizing precoder achieves a higher rate than the $\mathcal{I}_{\mathrm{RD}}^{\mathrm{FD}}$-maximizing precoder since the latter increases the interference at the FD relay's receiver due to the increased number of data streams transmitted by the relay.

Table I
Channel matrices used to generate the first three time slots in the figures.

| Slot number | Channel | Value |
|---|---|---|
| 1 | $\mathbf{H}_{SR}$ | $[0.013 + 0.0025i, 0.8374 - 0.8441i; 0.1166 - 0.3759i, 0.7537 + 0.2233i]$ |
| 1 | $\mathbf{H}_{RR}$ | $[1.6356 - 0.8668i, 0.1591 - 2.6461i; 0.7404 - 0.3748i, -0.7763 + 0.2951i]$ |
| 1 | $\mathbf{H}_{RD}$ | $[-0.2688 - 1.1046i, 1.0703 + 0.2583i; 0.8433 + 1.1624i, -0.3841 + 0.1363i]$ |
| 2 | $\mathbf{H}_{SR}$ | $[-0.3025 - 0.4487i, 0.6548 - 0.3400i; -0.4097 + 0.6069i, 0.0039 + 1.0534i]$ |
| 2 | $\mathbf{H}_{RR}$ | $[-0.445 + 0.9228i, -0.4446 + 0.5459i; -0.42 + 0.2586i, 0.2519 + 0.8876i]$ |
| 2 | $\mathbf{H}_{RD}$ | $[0.3088 - 1.7069i, 0.0019 + 0.2925i; -1.2754 + 0.2317i, -0.1195 - 0.4767i]$ |
| 3 | $\mathbf{H}_{SR}$ | $[0.184 - 1.0777i, 0.071 + 0.1647i; -0.3857 + 0.2473i, -0.5182 + 0.4624i]$ |
| 3 | $\mathbf{H}_{RR}$ | $[-0.5975 + 1.9031i, -0.347 - 0.4618i; -0.5693 + 0.2627i, -0.7111 + 0.3i]$ |
| 3 | $\mathbf{H}_{RD}$ | $[1.3800 + 1.5198i, 0.9294 + 1.6803i; -0.3835 + 0.5156i, 0.0726 + 0.5129i]$ |

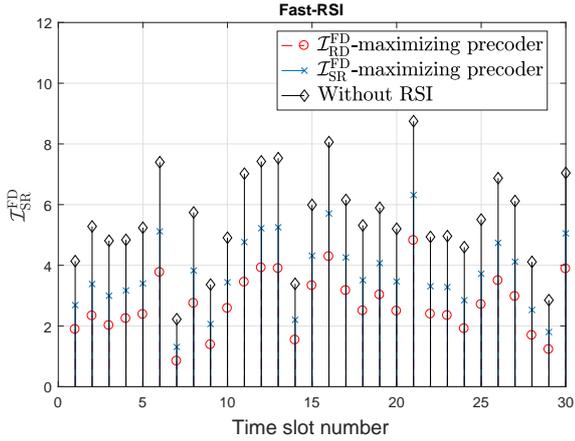

Figure 5. Achievable rate of source-relay link for the fast-RSI scenario.

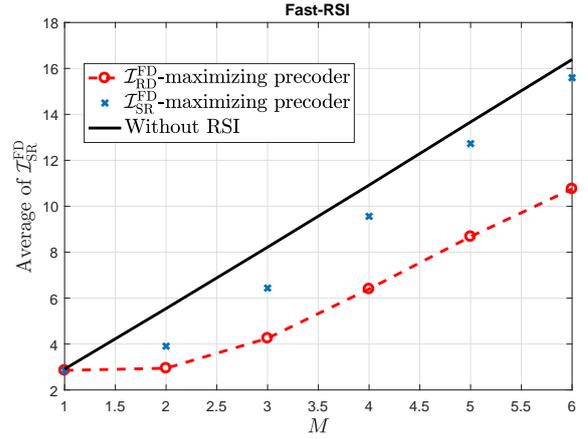

Figure 6. Average achievable rate of the source-relay link for the fast-RSI case under $\mathcal{I}_{\mathrm{RD}}^{\mathrm{FD}}$-maximizing and $\mathcal{I}_{\mathrm{SR}}^{\mathrm{FD}}$-maximizing precoders.

In Fig. 7, we show the minimum between the achievable rates of the source-relay and the relay-destination links. When $M = 2$ and for the given channel realizations, the $\mathcal{I}_{\mathrm{SR}}^{\mathrm{FD}}$-maximizing precoder outperforms the $\mathcal{I}_{\mathrm{RD}}^{\mathrm{FD}}$-maximizing precoder. However, this is not true in general since the $\mathcal{I}_{\mathrm{SR}}^{\mathrm{FD}}$-maximizing precoder degrades the achievable rate of the relay-destination link significantly, especially at high $M$. This is clear from the values of achievable rate evaluated for the other channel realizations as shown in Fig. 7 and in the average achievable rate curves presented in Fig. 8. It is noteworthy that when $M = 2$ as shown in Fig. 7, the relay might switch between $\mathcal{I}_{\mathrm{RD}}^{\mathrm{FD}}$-maximizing precoder and the $\mathcal{I}_{\mathrm{SR}}^{\mathrm{FD}}$-maximizing precoder to maximize the minimum achievable rate of the two hops, i.e., maximize $\min\{\mathcal{I}_{\mathrm{SR}}^{\mathrm{FD}}, \mathcal{I}_{\mathrm{RD}}^{\mathrm{FD}}\}$. As shown in Fig. 8, the expected value of the minimum between the achievable rate of the source-relay link and the achievable rate of the relay-destination link when the $\mathcal{I}_{\mathrm{SR}}^{\mathrm{FD}}$-maximizing precoder is slightly better than the $\mathcal{I}_{\mathrm{RD}}^{\mathrm{FD}}$-maximizing precoder when $M = 2$. Starting from $M = 3$, the $\mathcal{I}_{\mathrm{RD}}^{\mathrm{FD}}$-maximizing precoder is superior and can achieve very high rates. On the other hand, the $\mathcal{I}_{\mathrm{SR}}^{\mathrm{FD}}$-maximizing precoder remains fixed regardless of $M$ since the total achievable rate is determined by the minimum rate between the two communications links which is degraded by the use of a single data stream at the relay's transmit side.

### C. Fixed-Rate Transmission

In Fig. 9, we plot the throughput of our proposed scheme and the conventional FD scheme for $\mathcal{R} = 1$ bits/sec/Hz. In the conventional FD scheme, the source node and the relay cooperatively transmit the data in each time slot using the DF relaying scheme and the RSI is treated as a noise signal with a known variance. As shown in Fig. 9, the throughput increases by increasing the buffer size at the relay. This is expected since increasing the buffer size allows more data transfer to and from the relay; however, the increase is insignificant. Moreover, the throughput is fixed for all queue sizes that are greater than or equal to 3 packets. This implies that a data buffer with size 3 packets can be used without any throughput loss. Moreover, our proposed scheme achieves a throughput higher than that achieved by the conventional FD relaying. The throughput gain is more than $2866\%$ when the buffer's maximum size is $Q_{\max} \geq 3$ packets. We also plot an upper bound which is the case when the relay always has data packets and sends them to the destination. As shown in Fig. 9, the buffer-aided scheme outperforms the conventional FD scheme and it is closer to the upper bound. The throughput gap between the upper bound and the buffer-aided FD scheme is $6\%$ for $Q_{\max} \geq 2$ packets.

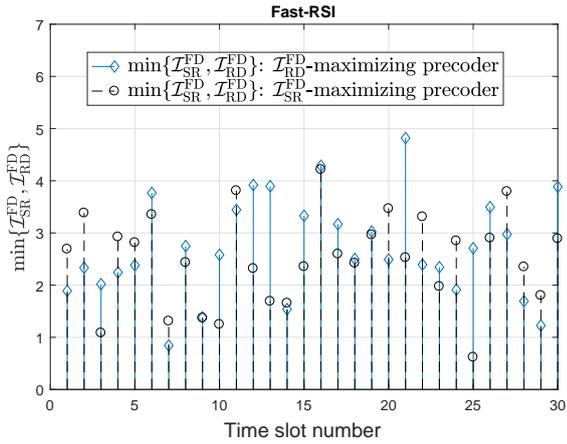

Figure 7. The minimum between the achievable rates of the source-relay link and the relay-destination link for the fast-RSI scenario. The case of $\mathcal{I}_{\mathrm{RD}}^{\mathrm{FD}}$-maximizing and $\mathcal{I}_{\mathrm{SR}}^{\mathrm{FD}}$-maximizing precoders are considered.

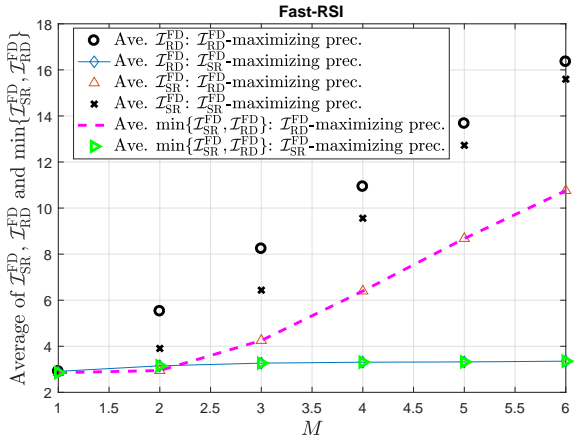

Figure 8. Average achievable rate of the source-relay link for the fast-RSI case under $\mathcal{I}_{\mathrm{RD}}^{\mathrm{FD}}$-maximizing and $\mathcal{I}_{\mathrm{SR}}^{\mathrm{FD}}$-maximizing precoders. The average of the minimum between the achievable rate of the source-relay and relay-destination links is plotted for both cases of $\mathcal{I}_{\mathrm{RD}}^{\mathrm{FD}}$-maximizing and $\mathcal{I}_{\mathrm{SR}}^{\mathrm{FD}}$-maximizing precoders.

In Fig. 10, we plot the throughput in bits/sec/Hz versus the transmission rate $\mathcal{R}$. The throughput in bits/sec/Hz is given by $\mu_{\mathrm{d}} \times \mathcal{R}$. The throughput in bits/sec/Hz increases with $\mathcal{R}$ until a peak is reached. This is expected since the throughput in packets/slot, given by $\mu_{\mathrm{d}}$, is monotonically non-increasing. Thus, multiplying $\mu_{\mathrm{d}}$ by $\mathcal{R}$ results in a peak at some $\mathcal{R}$. After that, the throughput decreases until it reaches zero. The value of $\mathcal{R}$ that maximizes the throughput for the buffer-aided FD case is $2.5$ bits/sec/Hz. The figure also shows the significant gain of our scheme relative to the conventional FD case. To show the impact of the RSI variance, we plotted the cases of $\sigma_{RR}^2 = 0$ dB and $\sigma_{RR}^2 = -10$ dB. The buffer-aided FD scheme does not depend on the RSI since it can be completely canceled as it was shown in the analytical proof in Appendix A and verified here through simulations. On the other hand, the conventional FD scheme suffers from self-interference and the throughput increases with decreasing RSI variance.

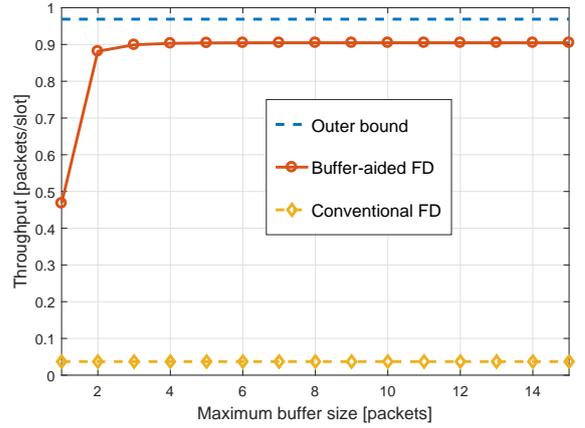

Figure 9. Throughput versus the maximum buffer size at the relay, $Q_{\max}$.

Finally, we demonstrate the impact of the number of antennas $M$ on the system's throughput in Fig. 11 for two different values of $\mathcal{R}$, i.e., $\mathcal{R} = 1$ and $\mathcal{R} = 6$ bits/sec/Hz. It can be seen that the throughput is monotonically nondecreasing with $M$. When $\mathcal{R} = 1$ bits/sec/Hz and $\mathcal{R} = 6$ bits/sec/Hz, the throughput is almost equal to 1 packet/slot, which is the maximum value for the system's throughput, for $M \geq 2$ and $M \geq 3$, respectively. Increasing $\mathcal{R}$ increases the outage probabilities of the communications' links and, hence, degrades the throughput measured in packets/slot.

## VII. CONCLUSIONS AND FUTURE WORK

We derived closed-form expressions for the achievable rates of the communications links in a buffered FD wireless relay network under the two scenarios of slow and fast RSI. We showed that, when the relay is equipped with a buffer, the impact of slow RSI can be completely eliminated in the time slots when the buffer is non-empty since the relay transmits a known codeword that is different from the source. That is, when the buffer is non-empty, the achievable rate of the source-relay link in the FD mode is equal to the achievable rate of the source-relay link without RSI. For fast RSI, we showed that the achievable rate of the source-relay link is degraded due to RSI and the degradation was quantified analytically. We designed two precoders that can be used at the relay, namely, the $\mathcal{I}_{\mathrm{SR}}^{\mathrm{FD}}$-maximizing and $\mathcal{I}_{\mathrm{RD}}^{\mathrm{FD}}$-maximizing precoders. We derived the closed-form expressions for the achievable rate of the source-relay and relay-destination links under each precoder. For the fixed-rate transmission scenarios, when the RSI is slow and the block size is large, we proposed an optimal scheme that maximizes the throughput, which is the number of packets received at the destination per time slot. Our numerical results showed that the throughput gain of our proposed buffer-aided FD scheme is substantial relative to the conventional FD scenario.

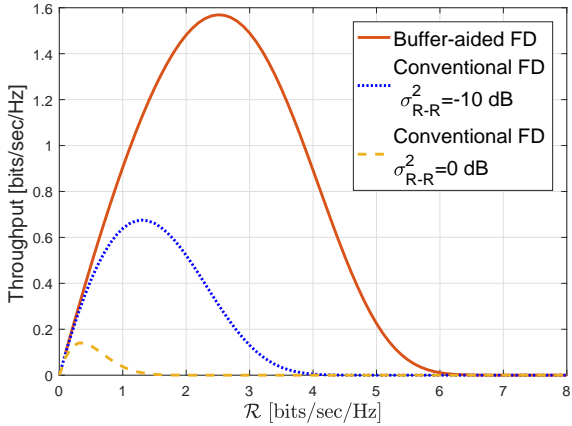

Figure 10. Throughput in bits/sec/Hz versus the transmission rate, $\mathcal{R}$.

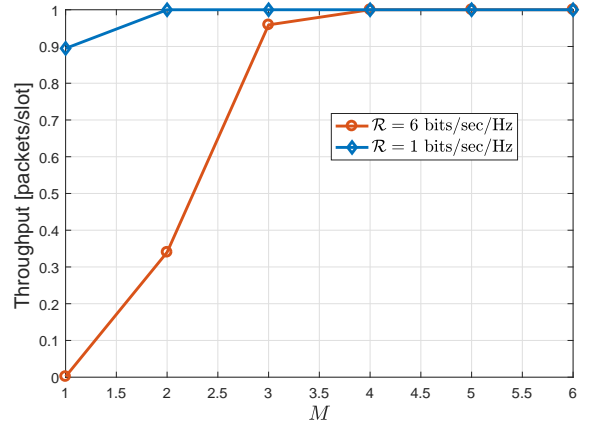

Figure 11. Throughput in packets/slot versus the number of antennas, $M$.

A possible future extension of this work is to consider the case of relay selection and study the gain of the buffers in such cases under FD constraints. Moreover, the multi-user scenario where multiple source nodes are communicating with a single or a set of relay nodes can be investigated.

## Appendix A
## Proof of Proposition 1

Starting from Eqn. (6), the Hermitian of the $\text{vec}\{\mathbf{Y}_R\}$ is given by

$$\begin{aligned} \text{vec}\{\mathbf{Y}_R\}^H = {} & \text{vec}\{\mathbf{X}_S\}^H \left(\mathbf{H}_{SR}^H \otimes \mathbf{I}_n\right) \\ & + \text{vec}\{\mathbf{H}_{RR}^\top\}^H \left(\mathbf{I}_M \otimes \boldsymbol{\Psi}^H \mathbf{X}_R^H\right) + \text{vec}\{\boldsymbol{\epsilon}_R\}^H. \end{aligned} \tag{33}$$

The expectation of $\text{vec}\{\mathbf{Y}_R\}\text{vec}\{\mathbf{Y}_R\}^H$ over $\mathbf{X}_S$ and $\mathbf{H}_{RR}$ is given by

$$\begin{aligned} \mathbb{E}\left\{\text{vec}\{\mathbf{Y}_R\}\text{vec}\{\mathbf{Y}_R\}^H\right\} = {} & \tilde{P}_S \left(\mathbf{H}_{SR} \otimes \mathbf{I}_n\right)\left(\mathbf{H}_{SR} \otimes \mathbf{I}_n\right)^H \\ & + \left(\mathbf{I}_M \otimes \mathbf{X}_R \boldsymbol{\Psi}\right) \boldsymbol{\Omega} \left(\mathbf{I}_M \otimes \mathbf{X}_R \boldsymbol{\Psi}\right)^H + \kappa_R \mathbf{I}_{nM}, \end{aligned} \tag{34}$$

where $\mathbb{E}\left\{\text{vec}\{\mathbf{X}_S\}\text{vec}\{\mathbf{X}_S\}^H\right\} = \tilde{P}_S \mathbf{I}_{nM}$ and $\boldsymbol{\Omega} = \mathbb{E}\{\text{vec}\{\mathbf{H}_{RR}^\top\}(\text{vec}\{\mathbf{H}_{RR}^\top\})^H\}$. Using the Kronecker product properties,

$$\begin{aligned} \mathbb{E}\left\{\text{vec}\{\mathbf{Y}_R\}\text{vec}\{\mathbf{Y}_R\}^H\right\} = {} & \tilde{P}_S \left(\mathbf{H}_{SR}\mathbf{H}_{SR}^H \otimes \mathbf{I}_n\right) \\ & + \left(\mathbf{I}_M \otimes \mathbf{X}_R \boldsymbol{\Psi}\right) \boldsymbol{\Omega} \left(\mathbf{I}_M \otimes \mathbf{X}_R \boldsymbol{\Psi}\right)^H + \kappa_R \mathbf{I}_{nM}. \end{aligned} \tag{35}$$

If $\boldsymbol{\Omega} = \sigma_{RR}^2 \mathbf{I}_{M^2}$, we can use the achievable rate expression for Gaussian vectors to obtain

$$\begin{aligned} \mathcal{I}_{SR}^{\text{FD}} = {} & \frac{1}{n} \log_2 \det \Big( \tilde{P}_S \left(\mathbf{H}_{SR}\mathbf{H}_{SR}^H \otimes \mathbf{I}_n\right) \\ & + \sigma_{RR}^2 \left(\mathbf{I}_M \otimes \mathbf{X}_R \boldsymbol{\Psi}\right) \left(\mathbf{I}_M \otimes \mathbf{X}_R \boldsymbol{\Psi}\right)^H + \kappa_R \mathbf{I}_{nM} \Big) \\ & - \frac{1}{n} \log_2 \det \left( \sigma_{RR}^2 \left(\mathbf{I}_M \otimes \mathbf{X}_R \boldsymbol{\Psi}\right) \left(\mathbf{I}_M \otimes \mathbf{X}_R \boldsymbol{\Psi}\right)^H + \kappa_R \mathbf{I}_{nM} \right). \end{aligned} \tag{36}$$

Consider the eigendecomposition for $\mathbf{H}_{SR}\mathbf{H}_{SR}^H = \mathbf{Q}_{SR}\boldsymbol{\Lambda}_{SR}\mathbf{Q}_{SR}^H$, where $\mathbf{Q}_{SR}$ is unitary and $\boldsymbol{\Lambda}_{SR}$ is diagonal. We can write the term $\left(\mathbf{H}_{SR}\mathbf{H}_{SR}^H \otimes \mathbf{I}_n\right)$ as $\left(\mathbf{Q}_{SR}\boldsymbol{\Lambda}_{SR}\mathbf{Q}_{SR}^H \otimes \mathbf{I}_n\right) = \left(\mathbf{Q}_{SR} \otimes \mathbf{I}_n\right)\left(\boldsymbol{\Lambda}_{SR} \otimes \mathbf{I}_n\right)\left(\mathbf{Q}_{SR}^H \otimes \mathbf{I}_n\right)$. Hence, the achievable rate expression can be rewritten as in (37) at the top of next page where $\eta_v$ denotes the $v$-th eigenvalue of $\mathbf{H}_{SR}\mathbf{H}_{SR}^H$.

## Appendix B
## Proof that the Optimal Precoder for slow-RSI is Rank-1

Let $\gamma_v = 1 + \frac{\tilde{P}_S \eta_v}{\kappa_R}$ and $\boldsymbol{\Gamma} = \sqrt{\frac{\sigma_{RR}^2}{\kappa_R}} \mathbf{X}_R$. Note that $\gamma_v > 0$. The achievable rate expression in (8) becomes

$$\begin{aligned} \mathcal{I}_{SR}^{\text{FD}} = {} & \frac{1}{n} \sum_{v=1}^{M} \log_2 \det\left(\gamma_v \mathbf{I}_n + \boldsymbol{\Gamma} \boldsymbol{\Psi} \boldsymbol{\Psi}^H \boldsymbol{\Gamma}^H\right) \\ & - \log_2 \det\left(\mathbf{I}_n + \boldsymbol{\Gamma} \boldsymbol{\Psi} \boldsymbol{\Psi}^H \boldsymbol{\Gamma}^H\right). \end{aligned} \tag{38}$$

Let $\boldsymbol{\Gamma} \boldsymbol{\Psi} \boldsymbol{\Psi}^H \boldsymbol{\Gamma}^H = \mathbf{Q}\boldsymbol{\Lambda}\mathbf{Q}^H = \mathbf{Q}\text{diag}\{\lambda_1, \lambda_2, \dots, \lambda_n\}\mathbf{Q}^H$, where $\lambda_1 \geq \lambda_2 \geq \dots \geq \lambda_n$ are the eigenvalues of $\boldsymbol{\Gamma} \boldsymbol{\Psi} \boldsymbol{\Psi}^H \boldsymbol{\Gamma}^H$ with $\sum_{k=1}^{n} \lambda_k = \text{Trace}\{\boldsymbol{\Gamma} \boldsymbol{\Psi} \boldsymbol{\Psi}^H \boldsymbol{\Gamma}^H\} = \frac{\sigma_{RR}^2 P_R}{\kappa_R} n = \lambda_{\text{tot}}$. The achievable rate is thus given by

$$\mathcal{I}_{SR}^{\text{FD}} = \frac{1}{n} \sum_{v=1}^{M} \sum_{k=1}^{n} \left[ \log_2\left(\gamma_v + \lambda_k\right) - \log_2\left(1 + \lambda_k\right) \right]. \tag{39}$$

By using $\lambda_1 = \lambda_{\text{tot}} - \sum_{k=2}^{n} \lambda_k$, the achievable rate can be rewritten as

$$\begin{aligned} \mathcal{I}_{SR}^{\text{FD}} = {} & \frac{1}{n} \sum_{v=1}^{M} \Bigg[ \log_2\left(\gamma_v + \lambda_{\text{tot}} - \sum_{k=2}^{n} \lambda_k\right) - \log_2\left(1 + \lambda_{\text{tot}} - \sum_{k=2}^{n} \lambda_k\right) \\ & + \sum_{k=2}^{n} \left(\log_2\left(\gamma_v + \lambda_k\right) - \frac{1}{n} \log_2\left(1 + \lambda_k\right)\right) \Bigg]. \end{aligned} \tag{40}$$

The derivative of $\mathcal{I}_{SR}^{\text{FD}}$ with respect to $\lambda_j$ ($j \in \{2, 3, \dots, n\}$), is given by (41) at the top of next page. If $\lambda_1 > \lambda_j$, then $\frac{\partial \mathcal{I}_{SR}^{\text{FD}}}{\partial \lambda_j} < 0$. Hence, $\lambda_j = 0$ maximizes the information rate. That is, $\lambda_1 = \lambda_{\text{tot}}$ and $\lambda_j = 0$ for $j \in \{2, 3, \dots, n\}$. If

$$
\begin{aligned}
\mathcal{I}_{SR}^{\mathrm{FD}} =& \frac{1}{n} \log_2 \det \Big( \tilde{P}_S \left( \boldsymbol{\Lambda}_{SR} \otimes \mathbf{I}_n \right) \\
& + \sigma_{RR}^2 \left( \mathbf{Q}_{SR}^H \otimes \mathbf{I}_n \right) \left( \mathbf{I}_M \otimes \mathbf{X}_R \boldsymbol{\Psi} \boldsymbol{\Psi}^H \mathbf{X}_R^H \right) \left( \mathbf{Q}_{SR} \otimes \mathbf{I}_n \right) + \kappa_R \mathbf{I}_{nM} \Big) \\
& - \frac{1}{n} \log_2 \det \left( \sigma_{RR}^2 \left( \mathbf{I}_M \otimes \mathbf{X}_R \boldsymbol{\Psi} \boldsymbol{\Psi}^H \mathbf{X}_R^H \right) + \kappa_R \mathbf{I}_{nM} \right) \\
=& \frac{1}{n} \log_2 \det \left( \tilde{P}_S \left( \boldsymbol{\Lambda}_{SR} \otimes \mathbf{I}_n \right) + \sigma_{RR}^2 \left( \mathbf{I}_M \otimes \mathbf{X}_R \boldsymbol{\Psi} \boldsymbol{\Psi}^H \mathbf{X}_R^H \right) + \kappa_R \mathbf{I}_{nM} \right) - \frac{M}{n} \log_2 \det \left( \sigma_{RR}^2 \mathbf{X}_R \boldsymbol{\Psi} \boldsymbol{\Psi}^H \mathbf{X}_R^H + \kappa_R \mathbf{I}_n \right) \\
=& \frac{1}{n} \sum_{v=1}^{M} \sum_{j=1}^{n} \left( \log_2 \left( 1 + \frac{\tilde{P}_S}{\kappa_R} \eta_v + \frac{\sigma_{RR}^2}{\kappa_R} \mathbf{X}_R(j) \boldsymbol{\Phi} \boldsymbol{\Phi}^H \mathbf{X}_R^H(j) \right) - \log_2 \left( 1 + \frac{\sigma_{RR}^2}{\kappa_R} \mathbf{X}_R(j) \boldsymbol{\Phi} \boldsymbol{\Phi}^H \mathbf{X}_R^H(j) \right) \right) \\
=& \frac{1}{n} \sum_{v=1}^{M} \sum_{j=1}^{n} \left( \log_2 \left( 1 + \frac{\tilde{P}_S}{\kappa_R} \eta_v + \frac{\sigma_{RR}^2}{\kappa_R} |\mathbf{X}_R(j)\mathbf{q}|^2 \right) - \log_2 \left( 1 + \frac{\sigma_{RR}^2}{\kappa_R} |\mathbf{X}_R(j)\mathbf{q}|^2 \right) \right).
\end{aligned}
\tag{37}
$$

$$
\begin{aligned}
\frac{\partial \mathcal{I}_{SR}^{\mathrm{FD}}}{\partial \lambda_j} =& \frac{1}{n \ln(2)} \sum_{v=1}^{M} \left[ -\frac{1}{\gamma + \lambda_{\mathrm{tot}} - \sum_{k=2}^{n} \lambda_k} + \frac{1}{1 + \lambda_{\mathrm{tot}} - \sum_{k=2}^{n} \lambda_k} + \frac{1}{\gamma + \lambda_j} - \frac{1}{1 + \lambda_j} \right] \\
=& \sum_{v=1}^{M} \frac{\gamma_v - 1}{n \ln(2)} \left[ \frac{1}{\left( \gamma_v + \lambda_{\mathrm{tot}} - \sum_{k=2}^{n} \lambda_k \right) \left( 1 + \lambda_{\mathrm{tot}} - \sum_{k=2}^{n} \lambda_k \right)} - \frac{1}{\left( \gamma_v + \lambda_j \right) \left( 1 + \lambda_j \right)} \right] \\
=& \sum_{v=1}^{M} \frac{\gamma_v - 1}{n \ln(2)} \left[ \frac{\gamma_v + \gamma_v \lambda_j + \lambda_j + \lambda_j^2 - \left( \gamma_v + \gamma_v \left( \lambda_{\mathrm{tot}} - \sum_{k=2}^{n} \lambda_k \right) + \left( \lambda_{\mathrm{tot}} - \sum_{k=2}^{n} \lambda_k \right) + \left( \lambda_{\mathrm{tot}} - \sum_{k=2}^{n} \lambda_k \right)^2 \right)}{\left( \gamma_v + \lambda_{\mathrm{tot}} - \sum_{k=2}^{n} \lambda_k \right) \left( 1 + \lambda_{\mathrm{tot}} - \sum_{k=2}^{n} \lambda_k \right) \left( \gamma_v + \lambda_j \right) \left( 1 + \lambda_j \right)} \right] \\
=& \sum_{v=1}^{M} \frac{\gamma_v - 1}{n \ln(2)} \left[ \frac{\gamma_v + \gamma_v \lambda_j + \lambda_j + \lambda_j^2 - \left( \gamma_v + \gamma_v \lambda_1 + \lambda_1 + \lambda_1^2 \right)}{\left( \gamma_v + \lambda_1 \right) \left( 1 + \lambda_1 \right) \left( \gamma_v + \lambda_j \right) \left( 1 + \lambda_j \right)} \right].
\end{aligned}
\tag{41}
$$

$\lambda_1 < \lambda_j$, $\frac{\partial \mathcal{I}_{SR}^{\mathrm{FD}}}{\partial \lambda_j} > 0$. In this case, $\lambda_j = \lambda_{\mathrm{tot}}$ and $\lambda_i = 0$ for all $i \neq j$ maximizes the information rate. Therefore, $\mathcal{I}_{SR}^{\mathrm{FD}}$ is maximized when one of the eigenvalues of $\boldsymbol{\Gamma} \boldsymbol{\Psi} \boldsymbol{\Psi}^H \boldsymbol{\Gamma}^H$ is $\lambda_{\mathrm{tot}}$ and the rest are zeros. This implies that the matrix $\boldsymbol{\Gamma} \boldsymbol{\Psi} \boldsymbol{\Psi}^H \boldsymbol{\Gamma}^H$ should be a rank-1 matrix. Since $\boldsymbol{\Gamma}$ has a rank of $M$, $\boldsymbol{\Psi} \boldsymbol{\Psi}^H$ and, consequently, $\boldsymbol{\Psi}$ is a rank-1 matrix.

## APPENDIX C
## ACHIEVABLE RATE OF THE RELAY-DESTINATION CHANNEL

The received signal matrix at the destination node is given by

$$
\mathbf{Y}_D = \mathbf{X}_R \boldsymbol{\Psi} \mathbf{H}_{RD}^\top + \boldsymbol{\epsilon}_D,
\tag{42}
$$

where $\mathbf{X}_R \in \mathbb{C}^{n \times M}$ is the data matrix transmitted by the relay node, $\mathbf{H}_{RD} \in \mathbb{C}^{M \times M}$ is the channel matrix between the relay node and the destination node, $\boldsymbol{\Psi} \in \mathbb{C}^{M \times M}$ is the data precoding matrix used at the relay node, and $\boldsymbol{\epsilon}_D \in \mathbb{C}^{N \times M}$ is the noise matrix at the destination node. Writing the matrix $\mathbf{Y}_D$ in a vector form, we have

$$
\mathrm{vec}\{\mathbf{Y}_D\} = (\mathbf{I}_M \otimes \mathbf{X}_R \boldsymbol{\Psi}) \, \mathrm{vec}\{\mathbf{H}_{RD}^\top\} + \mathrm{vec}\{\boldsymbol{\epsilon}_D\}.
\tag{43}
$$

The expectation of $\mathrm{vec}\{\mathbf{Y}_D\} \mathrm{vec}\{\mathbf{Y}_D\}^H$ over $\mathbf{X}_R$ is given by

$$
\begin{aligned}
\mathbb{E} \left\{ \mathrm{vec}\{\mathbf{Y}_D\} \mathrm{vec}\{\mathbf{Y}_D\}^H \right\} =& P_R \left( \mathbf{H}_{RD} \boldsymbol{\Psi}^\top \otimes \mathbf{I}_n \right) \left( \mathbf{H}_{RD} \boldsymbol{\Psi}^\top \otimes \mathbf{I}_n \right)^H \\
& + \kappa_D \mathbf{I}_{nM} \\
=& P_R \left( \mathbf{H}_{RD} \boldsymbol{\Psi}^\top \boldsymbol{\Psi}^* \mathbf{H}_{RD}^H \otimes \mathbf{I}_n \right) \\
& + \kappa_D \mathbf{I}_{nM}.
\end{aligned}
\tag{44}
$$

By using the achievable rate expression, we get

$$
\begin{aligned}
\mathcal{I}_{RD}^{\mathrm{FD}} =& \frac{1}{n} \log_2 \det \left( \mathbf{I}_{Mn} + \frac{P_R}{\kappa_D} \left( \mathbf{H}_{RD} \boldsymbol{\Psi}^\top \boldsymbol{\Psi}^* \mathbf{H}_{RD}^H \otimes \mathbf{I}_n \right) \right) \\
=& \log_2 \det \left( \mathbf{I}_M + \frac{P_R}{\kappa_D} \left( \mathbf{H}_{RD} \boldsymbol{\Psi}^\top \boldsymbol{\Psi}^* \mathbf{H}_{RD}^H \right) \right).
\end{aligned}
\tag{45}
$$

By using Sylvester's determinant identity, we have

$$
\mathcal{I}_{RD}^{\mathrm{FD}} = \log_2 \det \left( \mathbf{I}_M + \frac{P_R}{\kappa_D} \mathbf{H}_{RD}^H \mathbf{H}_{RD} \boldsymbol{\Psi}^\top \boldsymbol{\Psi}^* \right).
\tag{46}
$$

Consider the eigendecomposition $\mathbf{H}_{RD}^H \mathbf{H}_{RD} = \mathbf{Q}_{RD}^H \boldsymbol{\Lambda}_{RD} \mathbf{Q}_{RD}$. Thus, the achievable rate is given by

$$
\mathcal{I}_{RD}^{\mathrm{FD}} = \log_2 \det \left( \mathbf{I}_M + \frac{P_R}{\kappa_D} \boldsymbol{\Lambda}_{RD} \mathbf{Q}_{RD} \boldsymbol{\Psi}^\top \boldsymbol{\Psi}^* \mathbf{Q}_{RD}^H \right).
\tag{47}
$$

According to Hadamard's inequality for Hermitian positive semidefinite matrices, $\mathcal{I}_{RD}^{\mathrm{FD}}$ is maximized when $\mathbf{Q}_{RD} \boldsymbol{\Psi}^\top \boldsymbol{\Psi}^* \mathbf{Q}_{RD}^H$ is diagonal. Hence, $\boldsymbol{\Psi}^\top = \mathbf{Q}_{RD}^H \mathbf{E}$ where $\mathbf{E}$ is a diagonal matrix such that $\mathbf{E} \mathbf{E}^H$ contains the power fractions assigned to each data stream and its trace is equal to $\mathrm{Trace}\{\boldsymbol{\Psi} \boldsymbol{\Psi}^H\} = M$. Accordingly, the achievable rate of the relay-destination link in (47) is rewritten as

$$
\mathcal{I}_{RD}^{\mathrm{FD}} = \sum_{v=1}^{M} \log_2 \left( 1 + \frac{P_R}{\kappa_D} \boldsymbol{\Lambda}_{RD}(v) |E_v|^2 \right).
\tag{48}
$$

## APPENDIX D
## PROOF OF PROPOSITION 3

Setting $a = \frac{\sigma_{RR}^2}{\kappa_R}$, $b_v = \frac{\sigma_{RR}^2}{P_S \eta_v + \kappa_R} < a$, and $\alpha = \mathbf{q}^H \mathbf{X}_R^H \mathbf{X}_R \mathbf{q}$, we maximize the following function over $\alpha$

$$J(\alpha) = \sum_{v=1}^{M} \left( \log_2 \left(1 + b_v \alpha\right) - \log_2 \left(1 + a\alpha\right) \right). \quad (49)$$

Taking the first derivative with respect to $\alpha$, we get

$$\frac{\partial J(\alpha)}{\partial \alpha} = \frac{1}{\ln(2)} \sum_{v=1}^{M} \left( \frac{b_v}{1 + b_v \alpha} - \frac{a}{1 + a\alpha} \right)$$
$$= \frac{1}{\ln(2)} \frac{b_v - a}{(1 + b_v \alpha)(1 + a\alpha)}. \quad (50)$$

Since $b_v < a$, the derivative is always negative. Hence, $J(\alpha)$ is maximized when $\alpha$ is minimized. Now, $\alpha = \mathbf{q}^H \mathbf{X}_R^H \mathbf{X}_R \mathbf{q}$ is minimized when $\mathbf{q}$ is the normalized eigenvector corresponding to the minimum eigenvalue of $\mathbf{X}_R^H \mathbf{X}_R$. Substituting the precoder matrix, we get

$$\mathcal{I}_{SR}^{\mathrm{FD}} = \frac{1}{n} \sum_{v=1}^{M} \left( \log_2 \det \left( \tilde{P}_S \eta_v \mathbf{I}_n + \sigma_{RR}^2 \mathbf{X}_R \mathbf{q} \mathbf{q}^H \mathbf{X}_R^H + \kappa_R \mathbf{I}_n \right) \right.$$
$$\left. - \log_2 \det \left( \sigma_{RR}^2 \mathbf{X}_R \mathbf{q} \mathbf{q}^H \mathbf{X}_R^H + \kappa_R \mathbf{I}_n \right) \right)$$
$$= \sum_{v=1}^{M} \log_2 \left( 1 + \frac{\tilde{P}_S}{\kappa_R} \eta_v \right)$$
$$+ \frac{1}{n} \sum_{v=1}^{M} \left( \log_2 \left( 1 + \frac{\sigma_{RR}^2}{\tilde{P}_S \eta_v + \kappa_R} \mathbf{q}^H \mathbf{X}_R^H \mathbf{X}_R \mathbf{q} \right) \right.$$
$$\left. - \log_2 \left( 1 + \frac{\sigma_{RR}^2}{\kappa_R} \mathbf{q}^H \mathbf{X}_R^H \mathbf{X}_R \mathbf{q} \right) \right). \quad (51)$$

## APPENDIX E
## PROOF OF PROPOSITION 4

When the optimal precoder that maximizes the achievable rate of the relay-destination channel derived in Appendix C is used, $\boldsymbol{\Psi}$ is the product of a unitary and a diagonal matrix. Assuming that the matrix $\boldsymbol{\Psi} = \mathbf{E} \mathbf{Q}_{RD}^H$ is full rank, the achievable rate expression of the source-relay link can be rewritten as

$$\mathcal{I}_{SR}^{\mathrm{FD}} = \frac{1}{n} \sum_{v=1}^{M} \left( \log_2 \det \left( (\tilde{P}_S \eta_v + \kappa_R) \mathbf{I}_n + \sigma_{RR}^2 \mathbf{X}_R \mathbf{E} \mathbf{E}^H \mathbf{X}_R^H \right) \right.$$
$$\left. - \log_2 \det \left( \sigma_{RR}^2 \mathbf{X}_R \mathbf{E} \mathbf{E}^H \mathbf{X}_R^H + \kappa_R \mathbf{I}_n \right) \right). \quad (52)$$

Taking the data signal component as a common factor, we get

$$\mathcal{I}_{SR}^{\mathrm{FD}} = \frac{1}{n} \sum_{v=1}^{M} \left( n \log_2 \left( \tilde{P}_S \eta_v + \kappa_R \right) \right.$$
$$+ \log_2 \det \left( \mathbf{I}_n + \frac{\sigma_{RR}^2}{\tilde{P}_S \eta_v + \kappa_R} \mathbf{X}_R \mathbf{E} \mathbf{E}^H \mathbf{X}_R^H \right) \quad (53)$$
$$\left. - \log_2 \det \left( \sigma_{RR}^2 \mathbf{X}_R \mathbf{E} \mathbf{E}^H \mathbf{X}_R^H + \kappa_R \mathbf{I}_n \right) \right).$$

By using Sylvester's determinant identity, we get the expression in (10).

## APPENDIX F
## PROOF OF PROPOSITION 6

We vectorize the elements of the matrix $\mathbf{Y}_R$ in (15) to obtain

$$\mathrm{vec}\{\mathbf{Y}_R\} = (\mathbf{H}_{SR} \otimes \mathbf{I}_n) \, \mathrm{vec}\{\mathbf{X}_S\}$$
$$+ \left( \mathbf{I}_M \otimes \tilde{\mathbf{X}}_R \tilde{\boldsymbol{\Psi}} \right) \mathrm{vec}\{\tilde{\mathbf{H}}_{RR}\} + \mathrm{vec}\{\boldsymbol{\epsilon}_R\}. \quad (54)$$

The expected value of $\mathrm{vec}\{\mathbf{Y}_R\} \mathrm{vec}\{\mathbf{Y}_R\}^H$ is

$$\mathbb{E}\left\{ \mathrm{vec}\{\mathbf{Y}_R\} \mathrm{vec}\{\mathbf{Y}_R\}^H \right\} = \tilde{P}_S \left( \mathbf{H}_{SR} \otimes \mathbf{I}_n \right) \left( \mathbf{H}_{SR} \otimes \mathbf{I}_n \right)^H$$
$$+ \sigma_{RR}^2 \left( \mathbf{I}_M \otimes \tilde{\mathbf{X}}_R \tilde{\boldsymbol{\Psi}} \right) \left( \mathbf{I}_M \otimes \tilde{\mathbf{X}}_R \tilde{\boldsymbol{\Psi}} \right)^H$$
$$+ \kappa_R \mathbf{I}_{nM}, \quad (55)$$

by assuming $\mathrm{vec}\{\tilde{\mathbf{H}}_{RR}\} \mathrm{vec}\{\tilde{\mathbf{H}}_{RR}\}^H = \sigma_{RR}^2 \mathbf{I}_{nM}$. The achievable rate is thus given by

$$\mathcal{I}_{SR}^{\mathrm{FD}} = \frac{1}{n} \log_2 \det \left( \tilde{P}_S \left( \mathbf{H}_{SR} \mathbf{H}_{SR}^H \otimes \mathbf{I}_n \right) \right.$$
$$+ \sigma_{RR}^2 \left( \mathbf{I}_M \otimes \tilde{\mathbf{X}}_R \tilde{\boldsymbol{\Psi}} \right) \left( \mathbf{I}_M \otimes \tilde{\mathbf{X}}_R \tilde{\boldsymbol{\Psi}} \right)^H + \kappa_R \mathbf{I}_{nM} \right)$$
$$\left. - \frac{1}{n} \log_2 \det \left( \sigma_{RR}^2 \left( \mathbf{I}_M \otimes \tilde{\mathbf{X}}_R \tilde{\boldsymbol{\Psi}} \right) \left( \mathbf{I}_M \otimes \tilde{\mathbf{X}}_R \tilde{\boldsymbol{\Psi}} \right)^H + \kappa_R \mathbf{I}_{nM} \right). \quad (56)$$

Using the same matrix eigendecomposition for $\mathbf{H}_{SR} \mathbf{H}_{SR}^H$ as in the slow-RSI case, the achievable rate expression in (56) can be rewritten as:

$$\mathcal{I}_{SR}^{\mathrm{FD}} = \frac{1}{n} \log_2 \det \left( \tilde{P}_S \left( \boldsymbol{\Lambda}_{SR} \otimes \mathbf{I}_n \right) \right.$$
$$+ \sigma_{RR}^2 \left( \mathbf{Q}_{SR}^H \otimes \mathbf{I}_n \right) \left( \mathbf{I}_M \otimes \tilde{\mathbf{X}}_R \tilde{\boldsymbol{\Psi}} \left( \tilde{\mathbf{X}}_R \tilde{\boldsymbol{\Psi}} \right)^H \right) \left( \mathbf{Q}_{SR} \otimes \mathbf{I}_n \right)$$
$$+ \kappa_R \mathbf{I}_{nM} \right)$$
$$- \frac{1}{n} \log_2 \det \left( \left( \mathbf{I}_M \otimes \tilde{\mathbf{X}}_R \tilde{\boldsymbol{\Psi}} \left( \tilde{\mathbf{X}}_R \tilde{\boldsymbol{\Psi}} \right)^H \right) + \kappa_R \mathbf{I}_{nM} \right)$$
$$= \frac{1}{n} \log_2 \det \left( \tilde{P}_S \left( \boldsymbol{\Lambda}_{SR} \otimes \mathbf{I}_n \right) + \sigma_{RR}^2 \left( \mathbf{I}_M \otimes \tilde{\mathbf{X}}_R \tilde{\boldsymbol{\Psi}} \left( \tilde{\mathbf{X}}_R \tilde{\boldsymbol{\Psi}} \right)^H \right) \right.$$
$$+ \kappa_R \mathbf{I}_{nM} \right)$$
$$- \frac{M}{n} \log_2 \det \left( \sigma_{RR}^2 \tilde{\mathbf{X}}_R \tilde{\boldsymbol{\Psi}} \left( \tilde{\mathbf{X}}_R \tilde{\boldsymbol{\Psi}} \right)^H + \kappa_R \mathbf{I}_n \right). \quad (57)$$

After simplifications, we get the expression in (16).

## APPENDIX G
## PROOF OF PROPOSITION 7

By assuming the $\mathcal{I}_{SR}^{\mathrm{FD}}$-maximizing precoder of the slow-RSI case, which has the form $\boldsymbol{\Phi} = \sqrt{M} \mathbf{q} \mathbf{q}^H$, the information rate of the source-relay link is given by

$$\mathcal{I}_{SR}^{\mathrm{FD}} = \frac{1}{n} \sum_{v=1}^{M} \sum_{j=1}^{n} \left( \log_2 \left( 1 + \frac{\tilde{P}_S}{\kappa_R} \eta_v + \frac{\sigma_{RR}^2}{\kappa_R} \mathbf{X}_R(j) \boldsymbol{\Phi} \boldsymbol{\Phi}^H \mathbf{X}_R^H(j) \right) \right.$$
$$\left. - \log_2 \left( 1 + \frac{\sigma_{RR}^2}{\kappa_R} \mathbf{X}_R(j) \boldsymbol{\Phi} \boldsymbol{\Phi}^H \mathbf{X}_R^H(j) \right) \right)$$
$$= \frac{1}{n} \sum_{v=1}^{M} \sum_{j=1}^{n} \left( \log_2 \left( 1 + \frac{\tilde{P}_S}{\kappa_R} \eta_v + \frac{\sigma_{RR}^2}{\kappa_R} |\mathbf{X}_R(j) \mathbf{q}|^2 \right) \right.$$
$$\left. - \log_2 \left( 1 + \frac{\sigma_{RR}^2}{\kappa_R} |\mathbf{X}_R(j) \mathbf{q}|^2 \right) \right). \quad (58)$$

This can be simplified as follows

$$
\begin{aligned}
\mathcal{I}_{SR}^{\mathrm{FD}} = {} & \sum_{v=1}^{M} \log_2\left(1 + \frac{\tilde{P}_S}{\kappa_R}\eta_v\right) \\
& + \frac{1}{n}\sum_{v=1}^{M}\sum_{j=1}^{n}\Bigg(\log_2\left(1 + \frac{\frac{\sigma_{RR}^2}{\kappa_R}}{1 + \frac{\tilde{P}_S}{\kappa_R}\eta_v}|\mathbf{X}_R(j)\mathbf{q}|^2\right) \\
& - \log_2\left(1 + \frac{\sigma_{RR}^2}{\kappa_R}|\mathbf{X}_R(j)\mathbf{q}|^2\right)\Bigg).
\end{aligned} \tag{59}
$$

Since $\mathbf{q} \in \mathbb{C}^{M \times 1}$ is unit norm and $\mathbf{X}_R(j) \in \mathbb{C}^{1 \times M}$ is a complex Gaussian random vector with i.i.d. elements, $\mathbf{X}_R(j)\mathbf{q}$ is a Gaussian random variable with the same statistics as any element in $\mathbf{X}_R(j)$. From the strong law of large numbers, when $n$ is large, the term $\frac{1}{n}\sum_{v=1}^{M}\sum_{j=1}^{n}\left(\log_2\left(1 + \frac{\frac{\sigma_{RR}^2}{\kappa_R}}{1 + \frac{\tilde{P}_S}{\kappa_R}\eta_v}|\mathbf{X}_R(j)\mathbf{q}|^2\right) - \log_2\left(1 + \frac{\sigma_{RR}^2}{\kappa_R}|\mathbf{X}_R(j)\mathbf{q}|^2\right)\right)$ converges to its statistical mean. Accordingly, when $n \to \infty$, the achievable rate is given by (17).

## Appendix H
## Proof of Proposition 8

Assuming that the $\mathcal{I}_{\mathrm{RD}}^{\mathrm{FD}}$-maximizing precoder of the slow-RSI case, which has the form $\boldsymbol{\Phi} = \mathbf{E}\mathbf{Q}_{RD}^*$, is used by the relay, the information rate of the source-relay link in (61) for the fast-RSI case is rewritten as

$$
\begin{aligned}
\mathcal{I}_{SR}^{\mathrm{FD}} = {} & \frac{1}{n}\sum_{v=1}^{M}\sum_{j=1}^{n}\Bigg(\log_2\left(1 + \frac{\tilde{P}_S}{\kappa_R}\eta_v + \frac{\sigma_{RR}^2}{\kappa_R}\mathbf{X}_R(j)\mathbf{E}\mathbf{E}^H\mathbf{X}_R^H(j)\right) \\
& - \log_2\left(1 + \frac{\sigma_{RR}^2}{\kappa_R}\mathbf{X}_R(j)\mathbf{E}\mathbf{E}^H\mathbf{X}_R^H(j)\right)\Bigg).
\end{aligned} \tag{60}
$$

Assuming that $E_i$ is the $i$-th element on the main diagonal of $\mathbf{E}$, the information rate is thus given by

$$
\begin{aligned}
\mathcal{I}_{SR}^{\mathrm{FD}} = {} & \frac{1}{n}\sum_{v=1}^{M}\sum_{j=1}^{n}\Bigg(\log_2\left(\tilde{P}_S\eta_v + \sigma_{RR}^2\sum_{i=1}^{M}|X_{R,i}(j)|^2|E_i|^2 + \kappa_R\right) \\
& - \log_2\left(\sigma_{RR}^2\sum_{i=1}^{M}|X_{R,i}(j)|^2|E_i|^2 + \kappa_R\right)\Bigg) \\
= {} & \sum_{v=1}^{M}\log_2\left(1 + \frac{\tilde{P}_S}{\kappa_R}\eta_v\right) + \\
& + \frac{1}{n}\sum_{v=1}^{M}\sum_{j=1}^{n}\Bigg(\log_2\left(1 + \frac{\frac{\sigma_{RR}^2}{\kappa_R}\sum_{i=1}^{M}|X_{R,i}(j)|^2|E_i|^2}{1 + \frac{\tilde{P}_S}{\kappa_R}\eta_v}\right) \\
& - \log_2\left(1 + \frac{\sigma_{RR}^2}{\kappa_R}\sum_{i=1}^{M}|X_{R,i}(j)|^2|E_i|^2\right)\Bigg).
\end{aligned} \tag{61}
$$

From the strong law of large numbers, we can approximate $\log_2\left(\frac{\sigma_{RR}^2}{\kappa_R}\sum_{i=1}^{M}|X_{R,i}(j)|^2|E_i|^2 + 1\right)$ as follows

$$
\begin{aligned}
& M\lim_{n\to\infty}\sum_{j=1}^{n}\frac{\log_2\left(\frac{\sigma_{RR}^2}{\kappa_R}\sum_{i=1}^{M}|X_{R,i}(j)|^2 + 1\right)}{n} \\
& \to M\mathbb{E}\left\{\log_2\left(\frac{\sigma_{RR}^2}{\kappa_R}\mathcal{X}(j) + 1\right)\right\},
\end{aligned} \tag{62}
$$

where $\mathcal{X}(j) = \sum_{i=1}^{M}|X_{R,i}(j)|^2|E_i|^2$ is the weighted-sum of exponentially-distributed random variables and is equal in distribution to a scaled Chi-squared with $2M$ degrees of freedom when $|E_i|^2 = 1/M$ (i.e., the case of equal power allocation to the data streams). Note that $\mathbf{E}$ is assumed fixed over the entire codeword for a given channel realization and, hence, $\{\mathcal{X}(j)\}_{j=1}^{n}$ are i.i.d. random variables, which is true since all $\mathcal{X}(j)$ are identical.